\documentclass[aps,preprintnumbers,twocolumn,amsmath,amssymb,prl,floatfix,superscriptaddress]{revtex4-2}

\usepackage{graphicx}
\usepackage{dcolumn}
\usepackage{bm}
\usepackage{graphicx}
\usepackage{caption}
\usepackage{multirow}
\usepackage{subcaption}
\usepackage{ragged2e}
\usepackage{cancel}
\usepackage{comment}
\usepackage{xcolor}
\DeclareCaptionJustification{justified}{\justifying}
\usepackage{hyperref}
\usepackage{amssymb}
\usepackage{lipsum}

\begin{document}

\title{Thermodynamics and fractal Drude weights in the sine-Gordon model}

\author{Botond C. Nagy}\email{botond.nagy@edu.bme.hu}
\affiliation{Department of Theoretical Physics, Institute of Physics, Budapest University of Technology and Economics, H-1111 Budapest, M{\H u}egyetem rkp.~3.}
\affiliation{
BME-MTA Momentum Statistical Field Theory Research Group, Institute of Physics, Budapest University of Technology and Economics, H-1111 Budapest, M{\H u}egyetem rkp.~3.}
\author{M\'arton Kormos}\email{kormos.marton@ttk.bme.hu}
\affiliation{Department of Theoretical Physics, Institute of Physics, Budapest University of Technology and Economics, H-1111 Budapest, M{\H u}egyetem rkp.~3.}
\affiliation{MTA-BME Quantum Correlations Group (ELKH), Institute of Physics, Budapest University of Technology and Economics, H-1111 Budapest, M{\H u}egyetem rkp.~3.}
\affiliation{
BME-MTA Momentum Statistical Field Theory Research Group, Institute of Physics, Budapest University of Technology and Economics, H-1111 Budapest, M{\H u}egyetem rkp.~3.}
\author{G\'abor Tak\'acs}\email{takacs.gabor@ttk.bme.hu}
\affiliation{Department of Theoretical Physics, Institute of Physics, Budapest University of Technology and Economics, H-1111 Budapest, M{\H u}egyetem rkp.~3.}
\affiliation{MTA-BME Quantum Correlations Group (ELKH), Institute of Physics, Budapest University of Technology and Economics, H-1111 Budapest, M{\H u}egyetem rkp.~3.}
\affiliation{
BME-MTA Momentum Statistical Field Theory Research Group, Institute of Physics, Budapest University of Technology and Economics, H-1111 Budapest, M{\H u}egyetem rkp.~3.}

\date{24th May, 2023}

\begin{abstract}
The sine-Gordon model is a paradigmatic quantum field theory that provides the low-energy effective description of many gapped 1D systems. Despite this fact, its complete thermodynamic description in all its regimes has been lacking. Here we fill this gap and derive the framework that captures its thermodynamics and serves as the basis of its hydrodynamic description. As a first application, we compute the Drude weight characterising the ballistic transport of topological charge and demonstrate that its dependence on the value of the coupling shows a fractal structure, similar to the gapless phase of the XXZ spin chain. The thermodynamic framework can be applied to study other features of non-equilibrium dynamics in the sine-Gordon model using generalised hydrodynamics, opening the way to a wide array of theoretical studies and potential novel experimental predictions.  
\end{abstract}


\maketitle

\paragraph{Introduction and summary.---}
Transport properties such as electric conductivity provide fundamental insight into the dynamics of condensed matter \cite{transportbook}; however, their description is especially challenging  in the case of strongly correlated systems \cite{2016JSMTE..06.4010V}. In particular, integrable quantum many-body systems display many anomalous properties \cite{2021RvMP...93b5003B} due to ergodicity breaking captured by the Mazur inequality \cite{1969Phy....43..533M,1971Phy....51..277S}, and are primarily characterised by ballistic transport and finite Drude weights \cite{1995PhRvL..74..972C}. It has been studied quite intensively for the XXZ spin chain \cite{1997PhRvB..5511029Z,1999PhRvL..82.1764Z,2005JPSJ...74S.181B,2011PhRvB..84o5125H}, where a striking ``fractal" structure {of the spin Drude weight} was observed in the gapless phase \cite{2013PhRvL.111e7203P,2017PhRvL.119b0602I,2019PhRvL.122o0605L,2020PhRvB.101v4415A}, a phenomenon also known as ``popcorn" Drude weights \cite{2022JPhA...55X4005I}. Recently, the framework of generalised hydrodynamics (GHD) \cite{2016PhRvL.117t7201B,2016PhRvX...6d1065C,Doyon2019a,Essler2022,Bastianello2022} has led to a new approach to transport properties and, in particular, the computation of Drude weights  \cite{Doyon2017b,Ilievski2017b,2018PhRvB..97d5407B,2019ScPP....6....5U,2022JSMTE2022a4002D}. An experimental protocol to determine Drude weights was also proposed recently in  \cite{2017PhRvB..95f0406K}.

In this work, we consider ballistic transport of topological charge in the sine-Gordon model, which is a paradigmatic integrable quantum field theory with numerous applications to condensed matter physics \cite{tsvelik_2003,Giamarchi:743140} with the Hamilton operator
\begin{equation}
    H = \int \mathrm{d}x \left[\frac{1}{2} \left(\partial_t \phi\right)^2 + \frac{1}{2} \left(\partial_x \phi\right)^2 - \lambda \cos(\beta\phi)\right]\,,
\label{eq:sG_Ham}
\end{equation}
where $\phi(x)$ is a real scalar field, $\beta$ is the coupling, and $\lambda$ sets the mass scale. In particular, it is a model for a one-dimensional Mott insulator with applications to quasi-1D antiferromagnets, carbon nanotubes and organic conductors \cite{Controzzi2001,2005ffsc.book..684E}, {where electric conduction corresponds to transport of the topological charge carried by kink excitations}. It can also be realised in experiments with trapped ultra-cold atoms  \cite{2010PhRvL.105s0403C,2010Natur.466..597H,2017Natur.545..323S,2023arXiv230316221W}, which provide a well-controlled platform for studying non-equilibrium dynamics. In particular, the topological charge can be exploited in experiments to characterise the soliton dynamics \cite{2023arXiv230316221W}. {Other proposed realisations are via quantum circuits \cite{2021NuPhB.96815445R} or coupled spin chains \cite{Wybo2022}.}

\begin{figure*}
\centering
\captionsetup[subfigure]{justification=centering}
\begin{subfigure}{0.32\textwidth}
    \includegraphics[width=1\textwidth]{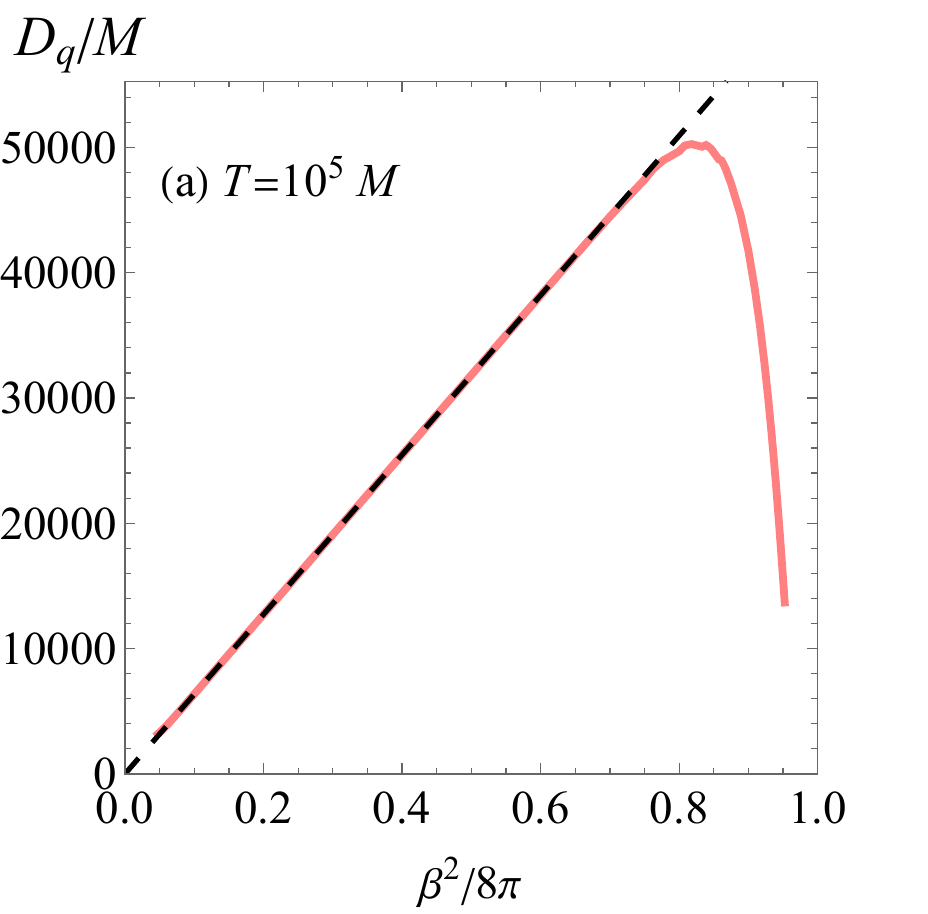}
    \end{subfigure}
    \hfill
\begin{subfigure}{0.32\textwidth}
    \includegraphics[width=1\textwidth]{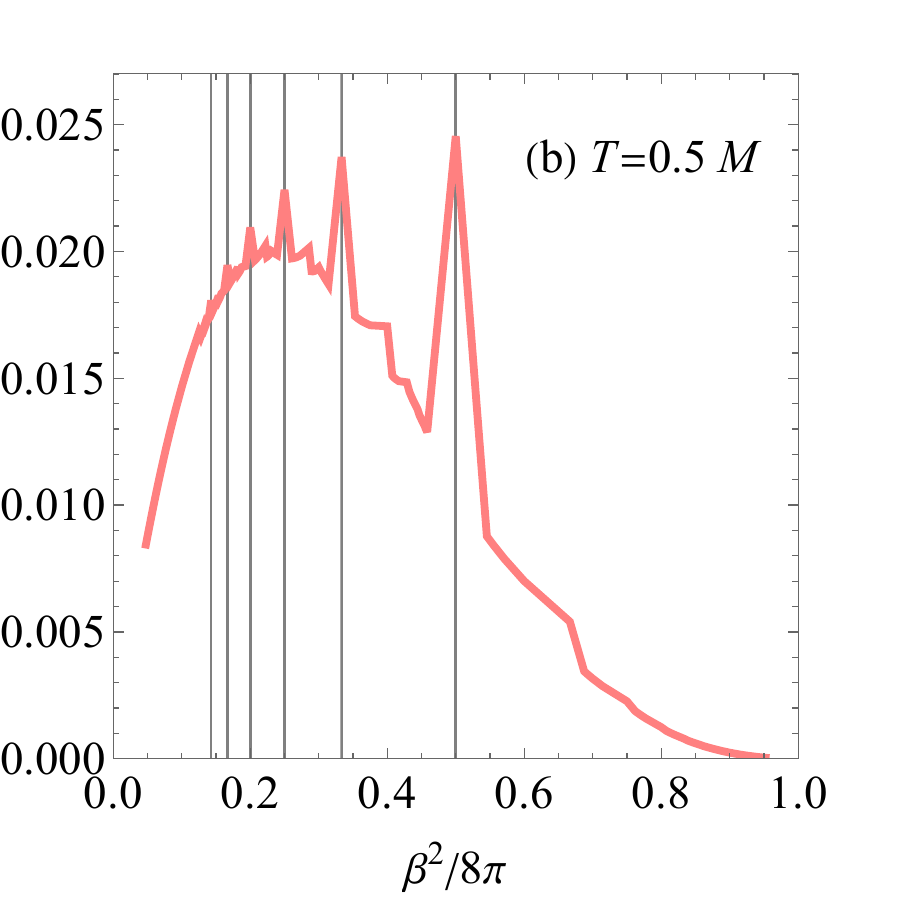}
\end{subfigure}
    \hfill
\begin{subfigure}{0.32\textwidth}
    \includegraphics[width=1\textwidth]{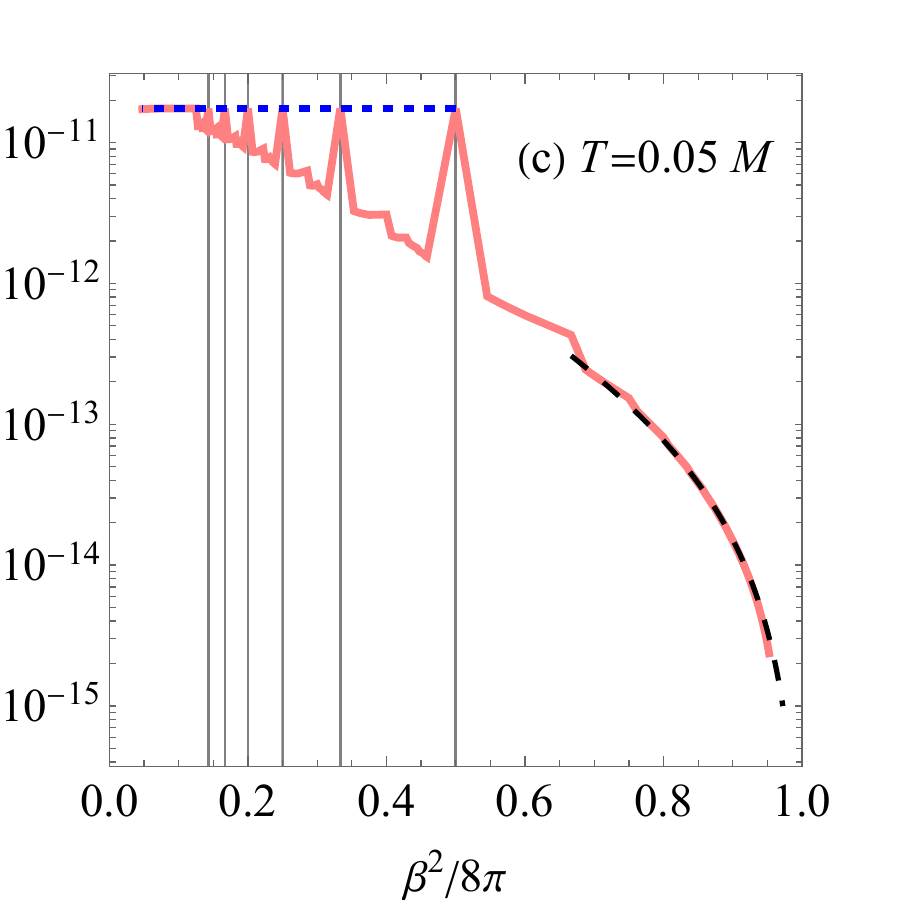}
\end{subfigure}
\caption{\label{fig:Drude} Drude weight in the sine-Gordon model calculated from Eq.(\ref{Drude_TBA}) as a function of the coupling strength $\beta^2/8\pi$ for (a) high; (b) moderate; and (c) low temperatures, at values of $\xi$ with at most two magnonic levels in the TBA system \eqref{TBA_eq}. The Drude weight was computed at discrete points which {are joined by} a red line in the plot to emphasise their discontinuous nature. At high temperatures, the fractal structure is washed out, while it gets more pronounced as the temperature is lower. We verified that the results from the bipartition protocol Eq. (\ref{Drude_bipartition}) agree with the TBA results within $0.1\%$ relative difference. 
The dashed line in (a) is the high-$T$ limit Eq.(\ref{high_T_Drude}), while in (c) the dotted blue line is the low-$T$ prediction for the reflectionless points \eqref{low_T_Drude_reflectionless}, and the black dashed line shows (\ref{low_T_Drude}). The vertical gridlines in (b) and (c) indicate the reflectionless points. Note also the logarithmic scale in (c).
}
\end{figure*}

The main stumbling block preventing progress in modelling the non-equilibrium dynamics in this important model has been the absence of an explicit thermodynamic description for general couplings, despite its known exact scattering theory \cite{1977CMaPh..55..183Z,1979AnPhy.120..253Z}. Thermodynamic Bethe Ansatz (TBA) systems for the sine-Gordon model have so far only been formulated explicitly for special values of the coupling \cite{ZAMOLODCHIKOV1991391,Tateo:1994pb,Bertini:2019lzy}, although a corresponding set of functional relations (the so-called $Y$ system) was conjectured for the general case in \cite{1995PhLB..355..157T}. Our first main result consists of the thermodynamic Bethe Ansatz (TBA) system for generic couplings with the corresponding dressing relations, which provides flexibility for applications to experiments by allowing generic values of the coupling parameter. We then exploit the TBA system to obtain the second main result: the Drude weight for charge transport in the sine-Gordon model. We demonstrate that the Drude weight considered as a function of the sine-Gordon coupling displays a fractal structure, as shown in Fig. \ref{fig:Drude}.

\paragraph{Thermodynamics of the sine-Gordon model.---}
The spectrum of the model is characterised in terms of the renormalised coupling constant $\xi=\beta^2/(8\pi-\beta^2).$ It contains a doublet of topologically charged kinks of mass $M$. In the attractive regime $0<\xi<1$, they also form $n_B=\lfloor1/\xi\rfloor$ different neutral bound states (breathers) of masses $m_{B_k}=2M\sin\frac{k\pi \xi}{2}$. The topological charge density and current are $\rho_q={\beta \partial_x \phi}/{2\pi}$ and $j_q=-{\beta \partial_t \phi}/{2\pi}$.

Thermodynamics at a given temperature $T$ and chemical potential $\mu$ conjugate to the topological charge can be computed using the thermodynamic Bethe Ansatz (TBA) \cite{yangyang1969,takahashi_1999,Zamolodchikov:1989cf}. For the so-called reflectionless points $1/\xi=n_B-1$ where all scattering is purely transmissive, it can be formulated in terms of the physical excitations (two kinks and $n_B$ breathers) \cite{Zamolodchikov:1989cf}. However, for generic couplings, it is given in terms of \emph{three} kinds of excitations: a single solitonic excitation $S$ 
accounting for the energy and momentum of the kinks, breathers $B_i$ ($i=1,\dots,n_B$);
and additional massless auxiliary excitations (``magnons'') which account for the topological charge carried by the kinks. Magnons can be classified by writing 
$\xi$ as a continued fraction
\begin{equation}
    \xi = \frac{1}{n_B+\displaystyle\frac{1}{\displaystyle\nu_1+\frac{1}{\nu_2+...}}} = \frac{1}{n_B+\displaystyle\frac{1}{\alpha}}\,,
\label{eq:contfraction}
\end{equation}
with $\nu_k$ magnon species at level $k$. We restrict our attention to cases with at most two magnonic levels, but extending to the general case is straightforward \cite{SM,inpreparation}.

\begin{figure*}[t!]
\centering
    \includegraphics[width=0.95\textwidth]{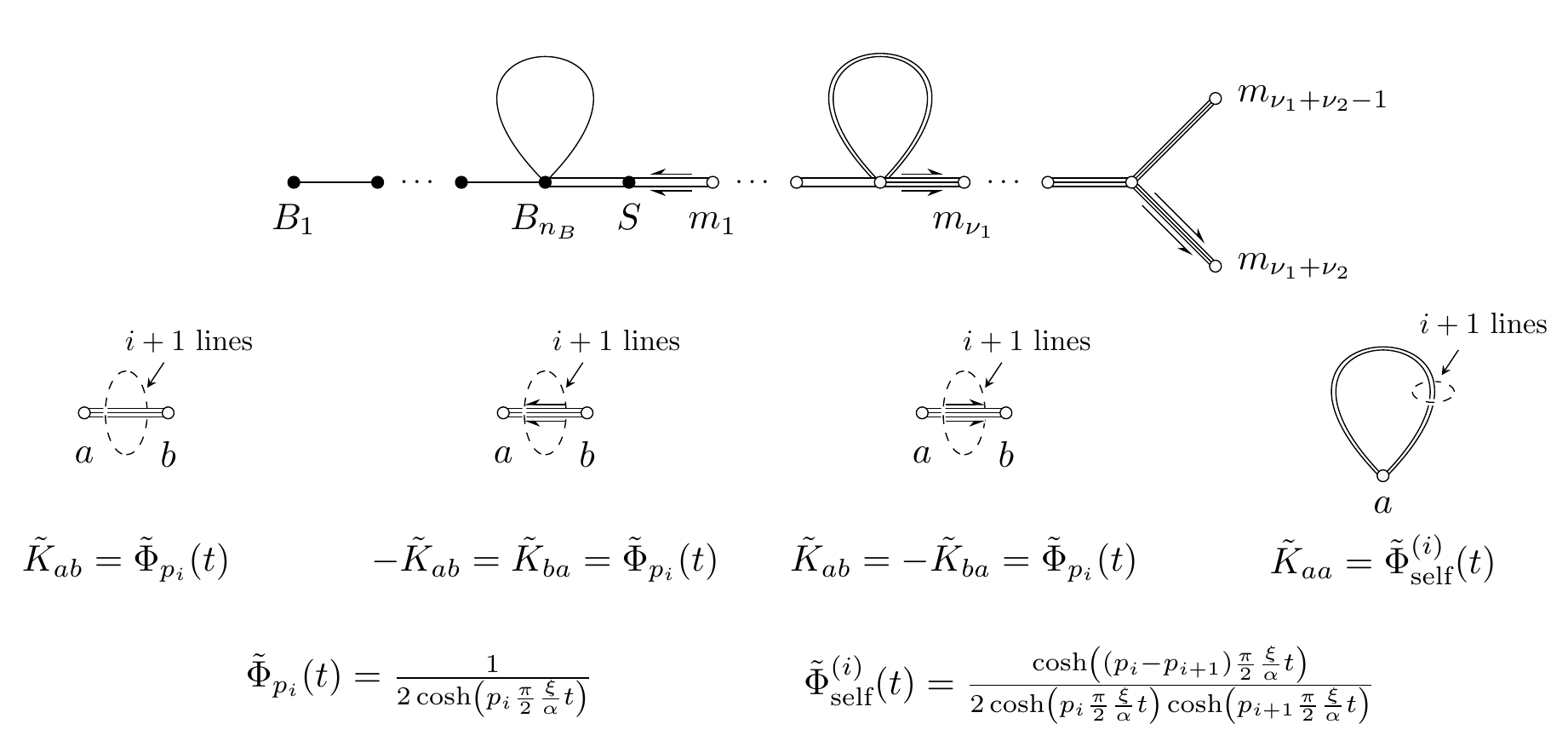}
    \caption{Diagram encoding the TBA kernels $K_{ij}(\theta)$ in terms of their Fourier transforms $\tilde{K}_{ab}(t)=\int {\mathrm{d}\theta}/{2\pi}\, K_{ab}(\theta) {e}^{-i\theta t}$  for the generic case with two magnonic levels (for the exceptions c.f. \cite{SM}). The diagram is similar to the one presented in \cite{boulat2019full} for the boundary sine-Gordon model. The kernel parameters are $p_0=\alpha=\nu_1+1/\nu_2$, $p_1=1$, $p_2=1/\nu_2$.}
    \label{kernel_diag}
\end{figure*}

Thermodynamic states corresponding to generalised Gibbs ensembles and containing an extensive number of particles can be characterised in terms of total and root densities $\rho_a^{\textrm{tot}}(\theta)$ and $\rho_a^{\textrm{r}}(\theta)$ describing the density of available and filled levels for excitation of type $a$ per unit rapidity $\theta$ and per unit volume. It is convenient to introduce the  so-called pseudo-energies $\epsilon_a(\theta)$ that parameterise the occupations as $\left[1+e^{\epsilon_a(\theta)}\right]^{-1}=\rho_a^\text{r}/\rho_a^\text{tot}$ and satisfy equations following from the minimisation of the free energy. We found that the TBA system of the sine-Gordon model can be written in the concise form
\begin{equation}
	\epsilon_a = w_a + \sum_b K_{ab} * \left( \sigma_b^{(1)}\epsilon_b - \sigma_b^{(2)} w_b + L_b \right)\,,
    \label{TBA_eq}
\end{equation}
where $L_a(\theta)=\log(1+{e}^{-\epsilon_a(\theta)})$ and the star denotes convolution.
The free energy $f$ density is given by 
\begin{equation}
    {f} = -T\sum_a \int \frac{\mathrm{d}\theta}{2\pi} M_a\cosh\theta L_a(\theta) \,.
\end{equation}
The kernels $K_{ab}(\theta)$ originate from the scattering of the excitations and can be encoded in a diagram \cite{ZAMOLODCHIKOV1991391,boulat2019full} as shown in Fig. \ref{kernel_diag}. The driving terms $w_a(\theta)$ carrying the dependence on the generalised chemical potentials (temperature $T$ and the chemical potential $\mu$ in thermal equilibrium), and the factors $\sigma_a^{(1,2)}$ are given in Table \ref{tab:TBA}.

The above TBA system, our first main result, can be validated by cross-checking its predictions for the free energy in (neutral) thermal states against the so-called Nonlinear Integral Equation \cite{1991JPhA...24.3111K,1995NuPhB.438..413D} as reported in \cite{SM}, and in the high-temperature limit ($\lambda=0$ in \eqref{eq:sG_Ham}) 
we recover the free energy of the free massless boson.

\begin{table*}
\centering
\bgroup
\def\arraystretch{1.2}%
\begin{tabular}{|l|l|c|c|c|c|c|}
\hline
Excitations                       & Labels                           & $w$                                          & $q$                        & $\eta$ & $\sigma^{(1)}$ & $\sigma^{(2)}$ \\ \hline\hline
Breathers                         & $B_i$, $i=1,...,n_B$             & ${M_{B_i}}\cosh(\theta)/{T} $              & $0$                        & $+1$   & $+1$           & $+1$           \\ \hline
Soliton                           & $S$                              & ${M} \cosh(\theta)/{T}$                    & $+1$                       & $+1$   & $0$            & $0$            \\ \hline
First level intermediate magnons  & $m_j$, $j=1,...,\nu_1-1$         & 0                                            & $-2\cdot j$                & $-1$   & $-1$           & $0$            \\ \hline
First level last magnon          & $m_{\nu_1}$                      & 0                                            & $-2$                       & $+1$   & $+1$           & $0$            \\ \hline
Second level intermediate magnons & $m_{\nu_1+k}$, $k=1,...,\nu_2-2$ & 0                                            & $-2\cdot(1+k \cdot \nu_1)$ & $+1$   & $+1$           & $0$            \\ \hline
Second level next-to-last magnon         & $m_{\nu_1+\nu_2-1}$, $(k=\nu_2-1)$           & $2(1+ \nu_1\cdot \nu_2)\cdot{\mu}/{T} $ & $-2\cdot(1+k \cdot \nu_1)$ & $+1$   & $+1$           & $0$            \\ \hline
Second level last magnon        & $m_{\nu_1+\nu_2}$                    & $ 2(1+ \nu_1\cdot \nu_2)\cdot{\mu}/{T}$ & $-2\cdot \nu_1$            & $-1$   & $0$            & $0$            \\ \hline
\end{tabular}
\egroup
\caption{Driving terms, topological charges and factors corresponding to the excitations in the TBA system \eqref{TBA_eq} and dressing equation \eqref{dressing} for two magnonic levels in the generic case (see \cite{SM} for other cases).\label{tab:TBA}}
\end{table*}

Conserved charges with values $Q_a(\theta)$ assigned to excitation $a$ with rapidity $\theta$ are also affected by the finite densities of excitations, and their {dressed values} $Q_a^{\textrm{dr}}$ can be obtained by solving the \emph{dressing equations} \cite{SM}
\begin{equation}
    \eta_a\,Q_a^{\text{dr}}\! =  Q_a + \sum_b K_{ab}*\left[\left(\sigma_b^{(1)}\!\!-\vartheta_b\right)\eta_b\,Q_b^{\text{dr}}\!-\sigma_b^{(2)} Q_b \right]
    \label{dressing}
\end{equation}
with $\vartheta_a(\theta)={\rho_a^{\textrm{r}}(\theta)}/{\rho_a^{\textrm{tot}}(\theta)}$ denoting the  \emph{filling fractions}. The total densities themselves can be obtained as $2\pi \rho^{\textrm{tot}}_a(\theta) = \left(\partial_{\theta}p_a\right)^{\textrm{dr}}\!\!(\theta)$, where  $p_a(\theta)=M_a\sinh\theta$ is the bare momentum of excitation type $a$ with rapidity $\theta$. The signs $\eta_a$ ensure the positivity of the densities and are given in Table \ref{tab:TBA}.
The effective velocity of excitations as a function of their rapidity $\theta$ can be obtained as $v^{\textrm{eff}}_a(\theta) =\left(\partial_{\theta}e_a\right)^{\textrm{dr}}\!\!(\theta)/\left(\partial_{\theta}p_a\right)^{\textrm{dr}}\!\!(\theta)$, where $e_a(\theta)=M_a\cosh\theta$ is the bare energy \cite{Doyon2019a,2020PhRvX..10a1054B,2020PhRvL.125g0602P}. 

\paragraph{Drude weight for the topological charge.---} 
Our TBA system can form the basis of the Generalised Hydrodynamics of the sine-Gordon model and can be used to compute various physical quantities in and out of equilibrium. We focus on the Drude weight of the topological charge defined from the connected current correlator as 
\begin{equation}
D_q={\lim_{\tau\to\infty}} \frac{1}{2\tau}\int_{-\tau}^{\tau} \mathrm{d}t \int \mathrm{d}x \,\langle j_q(x,t) j_q(0,0)\rangle_c
\end{equation}
{which can be expressed from the TBA as} \cite{Doyon2017b,Ilievski2017b,2022JSMTE2022a4002D}
\begin{equation}
    D_q = \sum_a\!\int\!\mathrm{d}\theta \rho_a^{\textrm{tot}}(\theta) \vartheta_a(\theta)[1-\vartheta_a(\theta)]\left[v_a^{\textrm{eff}}(\theta)q_a^{\textrm{dr}}(\theta)\right]^2
    \label{Drude_TBA}
\end{equation}
where $q_a^{\textrm{dr}}(\theta)$ is the dressed topological charge of excitation $a$ with rapidity $\theta$. Solving the TBA system \eqref{TBA_eq} for different values of $T$ and $\xi$ (with $\mu=0$), evaluating the dressed quantities using \eqref{dressing} and substituting them into \eqref{Drude_TBA} results in the data shown in Fig. \ref{fig:Drude}.

Another way to obtain the Drude weights is to consider a bipartitioned initial state with a small chemical potential difference $\delta\mu$ between the half-systems $x>0$ and $x<0$. Initial densities on the two sides can be obtained from the dressing equation \eqref{dressing} using the appropriate solution of the TBA system \eqref{TBA_eq} with finite temperature $T$ and chemical potentials $\mu=\pm\delta\mu/2$, and the subsequent time evolution can be computed by a simple application of the GHD as described in \cite{SM}. The asymptotic current profile is a limit along ``rays" $\textrm{j}\left(\zeta\right)=\lim_{t\rightarrow\infty} j(x=\zeta t, t)$, and the Drude weight is  \cite{2017PhRvL.119b0602I,2018PhRvB..97d5407B}
\begin{equation}
    D_q = \left.\frac{\partial}{\partial \delta \mu} \int \mathrm{d}\zeta\ \textrm{j}\left(\zeta\right)\right|_{\delta\mu=0}\,.
    \label{Drude_bipartition}
\end{equation}
We verified that this method gives the same result as Eq. \eqref{Drude_TBA} within the attainable numerical precision.

\paragraph{Special cases.---} 
Besides numerical results, we can also obtain analytic expressions in appropriate limits. For high temperatures $T/M\gg 1$, the fractal structure is suppressed and the Drude weight is a continuous function of the renormalised coupling $\xi$ \cite{SM}:
\begin{equation}
    D^{\textrm{high-}T}_q = \frac{2 T}{\pi} \frac{\xi}{\xi + 1}= \frac{T\beta^2}{4\pi^2}\, .
    \label{high_T_Drude}
\end{equation}
In this regime, the dynamics is independent of the coupling $\lambda$ in \eqref{eq:sG_Ham} and is described by a free massless boson, which reproduces \eqref{high_T_Drude} by an easy derivation \cite{SM}. 

In the opposite limit of low temperature $T/M\ll 1$ and for a single magnonic level in the repulsive regime (i.e. $\xi=2,3,\dots$), the analytic result is \cite{SM}
{
\begin{equation}
    D^{\textrm{low-}T}_q =  e^{-M/T}\frac{\sqrt{2}T^{5/2} }{\pi^{3/2}M^{3/2} }
    \frac{2\pi/\xi - \sin(2\pi/\xi)}{\xi\sin^2(\pi/\xi)}\,.
    \label{low_T_Drude}
\end{equation}
}
{In the attractive regime, at reflectionless couplings $\xi=1/(n_B-1)$, we find 
in the low-temperature limit}
\begin{equation}
    D^{\textrm{low-}T}_q\! = 2 \int \frac{\mathrm{d}\theta}{2\pi}  \frac{M\cosh\theta\,{e}^{-M \cosh\theta/T}}{\left(1\!+\!{e}^{-M\cosh\theta/T }\right)^2} \tanh^2\theta 
    \label{low_T_Drude_reflectionless}
\end{equation}
{independent of the coupling. This} result can be obtained simply by considering the kinks/antikinks as non-interacting fermions with energy $M\cosh\theta$ and velocity $\tanh\theta$, since in the absence of the massless magnons all excitations are massive, and therefore the effects of the scattering embodied in the integral terms of Eqs. (\ref{TBA_eq},\ref{dressing}) are exponentially suppressed \cite{SM}.  

As shown in Fig. \ref{fig:Drude}, these analytic limiting cases agree very well with the numerical values computed from \eqref{Drude_TBA}.

\paragraph{Discussion and outlook.---}
{We have demonstrated the finiteness of the Drude weight of the topological charge in the sine-Gordon model, which implies the existence of yet unknown conserved quantities that are odd under charge reflection. Moreover, it has a fractal structure {(as argued recently in \cite{2022JPhA...55X4005I})}, which is the only appearance of such a commensurability effect so far besides the prototypical XXZ spin chain.}

\begin{figure}
    \centering
    \includegraphics[width=0.47\textwidth]{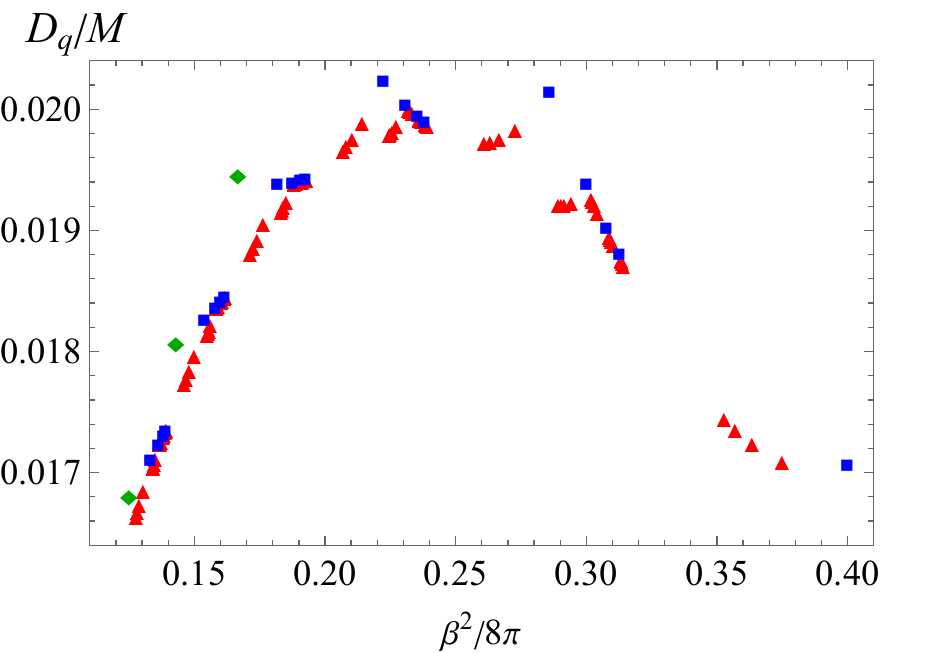}
    \caption{A close-up view of the fractal structure of Drude weights from Fig. \ref{fig:Drude} (b) ($T\!=\!M/2$) with separate markings for different numbers of magnon levels. Green diamonds: reflectionless couplings ($N=0$, no magnons), blue squares: $N=1$ magnonic level, red triangles: $N=2$ magnonic levels.}
    \label{fig:fractal_structure}
\end{figure}

The Drude weight (c.f. Fig. \ref{fig:Drude}) has some interesting properties. First, we note that $D_q$ approaches zero at the Kosterlitz-Thouless point $\beta^2/8\pi=1$, as demonstrated by the numerical data and also by the explicit expression \eqref{low_T_Drude} for low temperatures. {Note that the limits $\xi\to\infty$ and $T\to\infty$ do not commute.} Second, the values of $D_q$ for a fixed number $N$ of magnonic levels appear to form regular sequences when considered as a function of the deepest level integer $\nu_N$ in the continued fraction expansion \eqref{eq:contfraction} with all other (lower level) integers kept fixed, as demonstrated in Fig. \ref{fig:fractal_structure}. Third, increasing the depth of the continued fraction \eqref{eq:contfraction}, i.e., the number of magnonic levels $N$ suppresses the fractal structure, suggesting that for irrational values of the coupling $\xi$ corresponding to infinite continued fractions, the values of the Drude weight $D_q$ lie on a limiting envelope curve. In fact, the Drude weight $D_q$ for an irrational $\xi$ is expected to be successively approximated by truncating the continued fraction expansion \eqref{eq:contfraction} progressively deeper, i.e., with an increasing number of magnonic levels \cite{2020PhRvB.101v4415A}.

In addition to the new results for the Drude weights, the TBA system \eqref{TBA_eq} and the associated dressing equations \eqref{dressing} open the way to new applications of generalised hydrodynamics to the sine-Gordon model 
at generic values of the coupling. We demonstrated the application to non-equilibrium dynamics by {studying a partitioning protocol \cite{SM}, but more general inhomogeneous setups can be studied providing new predictions relevant to condensed matter and cold atom experiments. It is especially interesting in view of studies of generalised hydrodynamics in atom chip experiments \cite{2019PhRvL.122i0601S}, given the realisation of the sine-Gordon model in this setting \cite{2017Natur.545..323S}. Phase imprinting and the ability of designing arbitrary space-dependent optical potentials in coupled atomic condensates allows for the study of controlled non-equilibrium situations \cite{2018PhRvL.120q3601P, 2019OExpr..2733474T}.}
{Apart from the topological charge, transport of other quantities, such as the energy, can also be studied. Fluctuations, full counting statistics, and dynamical correlations of vertex operators can also be described via the ballistic fluctuation theory \cite{Doyon2019b,Myers2020}, extending the recent study \cite{DelVecchio2023}, which we expect to provide another manifestation of the fractal structure. Finally, it would be very interesting to analyse the diffusive corrections to the ballistic behaviour and the possibility of superdiffusion \cite{Bulchandani2021} which we plan to address in a forthcoming publication.}

\begin{acknowledgments}
\paragraph{Acknowledgements.---} {We thank S. Erne and J. Schmiedmayer for discussions, and also A. Bastianello, B. Doyon, E. Ilievski and B. Pozsgay for their comments on a preliminary version of this manuscript.} This work was supported by the  National Research, Development and Innovation Office (NKFIH) through the OTKA Grant ANN 142584. BN was also partially supported by the National Research Development and Innovation Office of Hungary via the scholarship ÚNKP-22-3-I-BME-12, while GT was partially supported by the Quantum Information National Laboratory of Hungary (Grant No. 2022-2.1.1-NL-2022-00004). 
\end{acknowledgments}

\bibliographystyle{utphys}
\bibliography{drude}

\providecommand{\href}[2]{#2}\begingroup\raggedright\begin{thebibliography}{10}

\bibitem{transportbook}
H.~Bruus and K.~Flensberg, {\em Many-body quantum theory in condensed matter
  physics: an introduction}.
\newblock Oxford University Press, United Kingdom, 2003.

\bibitem{2016JSMTE..06.4010V}
R.~{Vasseur} and J.~E. {Moore}, ``{Nonequilibrium quantum dynamics and
  transport: from integrability to many-body localization},''
  \href{http://dx.doi.org/10.1088/1742-5468/2016/06/064010}{{\em J. Stat. Mech.
  Theor. Exp.} {\bfseries 6} (2016) 064010},
  \href{http://arxiv.org/abs/1603.06618}{{\ttfamily arXiv:1603.06618
  [cond-mat.str-el]}}.

\bibitem{2021RvMP...93b5003B}
B.~{Bertini}, F.~{Heidrich-Meisner}, C.~{Karrasch}, T.~{Prosen},
  R.~{Steinigeweg}, and M.~{{\v{Z}}nidari{\v{c}}}, ``{Finite-temperature
  transport in one-dimensional quantum lattice models},''
  \href{http://dx.doi.org/10.1103/RevModPhys.93.025003}{{\em Rev. Mod. Phys.}
  {\bfseries 93} (2021) 025003},
  \href{http://arxiv.org/abs/2003.03334}{{\ttfamily arXiv:2003.03334
  [cond-mat.str-el]}}.

\bibitem{1969Phy....43..533M}
P.~{Mazur}, ``{Non-ergodicity of phase functions in certain systems},''
  \href{http://dx.doi.org/10.1016/0031-8914(69)90185-2}{{\em Physica}
  {\bfseries 43} (1969) 533--545}.

\bibitem{1971Phy....51..277S}
M.~{Suzuki}, ``{Ergodicity, constants of motion, and bounds for
  susceptibilities},''
  \href{http://dx.doi.org/10.1016/0031-8914(71)90226-6}{{\em Physica}
  {\bfseries 51} (1971) 277--291}.

\bibitem{1995PhRvL..74..972C}
H.~{Castella}, X.~{Zotos}, and P.~{Prelov{\v{s}}ek}, ``{Integrability and Ideal
  Conductance at Finite Temperatures},''
  \href{http://dx.doi.org/10.1103/PhysRevLett.74.972}{{\em Phys. Rev. Lett.}
  {\bfseries 74} (1995) 972--975},
  \href{http://arxiv.org/abs/cond-mat/9411005}{{\ttfamily
  arXiv:cond-mat/9411005 [cond-mat]}}.

\bibitem{1997PhRvB..5511029Z}
X.~{Zotos}, F.~{Naef}, and P.~{Prelovsek}, ``{Transport and conservation
  laws},'' \href{http://dx.doi.org/10.1103/PhysRevB.55.11029}{{\em Phys. Rev.
  B} {\bfseries 55} (1997) 11029--11032},
  \href{http://arxiv.org/abs/cond-mat/9611007}{{\ttfamily
  arXiv:cond-mat/9611007 [cond-mat.str-el]}}.

\bibitem{1999PhRvL..82.1764Z}
X.~{Zotos}, ``{Finite Temperature Drude Weight of the One-Dimensional Spin- 1/2
  Heisenberg Model},''
  \href{http://dx.doi.org/10.1103/PhysRevLett.82.1764}{{\em Phys. Rev. Lett.}
  {\bfseries 82} (1999) 1764--1767},
  \href{http://arxiv.org/abs/cond-mat/9811013}{{\ttfamily
  arXiv:cond-mat/9811013 [cond-mat]}}.

\bibitem{2005JPSJ...74S.181B}
J.~{Benz}, T.~{Fukui}, A.~{Kl{\"u}mper}, and C.~{Scheeren}, ``{On the Finite
  Temperature Drude Weight of the Anisotropic Heisenberg Chain},''
  \href{http://dx.doi.org/10.1143/JPSJS.74S.181}{{\em J. Phys. Soc. Jpn}
  {\bfseries 74} (2005) 181--190},
  \href{http://arxiv.org/abs/cond-mat/0502516}{{\ttfamily
  arXiv:cond-mat/0502516 [cond-mat.str-el]}}.

\bibitem{2011PhRvB..84o5125H}
J.~{Herbrych}, P.~{Prelov{\v{s}}ek}, and X.~{Zotos}, ``{Finite-temperature
  Drude weight within the anisotropic Heisenberg chain},''
  \href{http://dx.doi.org/10.1103/PhysRevB.84.155125}{{\em Phys. Rev. B}
  {\bfseries 84} (2011) 155125},
  \href{http://arxiv.org/abs/1107.3027}{{\ttfamily arXiv:1107.3027
  [cond-mat.str-el]}}.

\bibitem{2013PhRvL.111e7203P}
T.~{Prosen} and E.~{Ilievski}, ``{Families of Quasilocal Conservation Laws and
  Quantum Spin Transport},''
  \href{http://dx.doi.org/10.1103/PhysRevLett.111.057203}{{\em Phys. Rev.
  Lett.} {\bfseries 111} (2013) 057203},
  \href{http://arxiv.org/abs/1306.4498}{{\ttfamily arXiv:1306.4498
  [cond-mat.stat-mech]}}.

\bibitem{2017PhRvL.119b0602I}
E.~{Ilievski} and J.~{De Nardis}, ``{Microscopic Origin of Ideal Conductivity
  in Integrable Quantum Models},''
  \href{http://dx.doi.org/10.1103/PhysRevLett.119.020602}{{\em Phys. Rev.
  Lett.} {\bfseries 119} (2017) 020602},
  \href{http://arxiv.org/abs/1702.02930}{{\ttfamily arXiv:1702.02930
  [cond-mat.stat-mech]}}.

\bibitem{2019PhRvL.122o0605L}
M.~{Ljubotina}, L.~{Zadnik}, and T.~{Prosen}, ``{Ballistic Spin Transport in a
  Periodically Driven Integrable Quantum System},''
  \href{http://dx.doi.org/10.1103/PhysRevLett.122.150605}{{\em Phys. Rev.
  Lett.} {\bfseries 122} (2019) 150605},
  \href{http://arxiv.org/abs/1901.05398}{{\ttfamily arXiv:1901.05398
  [cond-mat.stat-mech]}}.

\bibitem{2020PhRvB.101v4415A}
U.~{Agrawal}, S.~{Gopalakrishnan}, R.~{Vasseur}, and B.~{Ware}, ``{Anomalous
  low-frequency conductivity in easy-plane XXZ spin chains},''
  \href{http://dx.doi.org/10.1103/PhysRevB.101.224415}{{\em Phys. Rev. B}
  {\bfseries 101} (2020) 224415},
  \href{http://arxiv.org/abs/1909.05263}{{\ttfamily arXiv:1909.05263
  [cond-mat.stat-mech]}}.

\bibitem{2022JPhA...55X4005I}
E.~{Ilievski}, ``{Popcorn Drude weights from quantum symmetry},''
  \href{http://dx.doi.org/10.1088/1751-8121/acaa77}{{\em J. Phys. A Math. Gen.}
  {\bfseries 55} (2022) 504005},
  \href{http://arxiv.org/abs/2208.01446}{{\ttfamily arXiv:2208.01446
  [cond-mat.stat-mech]}}.

\bibitem{2016PhRvL.117t7201B}
B.~{Bertini}, M.~{Collura}, J.~{De Nardis}, and M.~{Fagotti}, ``{Transport in
  Out-of-Equilibrium XXZ Chains: Exact Profiles of Charges and Currents},''
  \href{http://dx.doi.org/10.1103/PhysRevLett.117.207201}{{\em Phys. Rev.
  Lett.} {\bfseries 117} (2016) 207201},
  \href{http://arxiv.org/abs/1605.09790}{{\ttfamily arXiv:1605.09790
  [cond-mat.stat-mech]}}.

\bibitem{2016PhRvX...6d1065C}
O.~A. {Castro-Alvaredo}, B.~{Doyon}, and T.~{Yoshimura}, ``{Emergent
  Hydrodynamics in Integrable Quantum Systems Out of Equilibrium},''
  \href{http://dx.doi.org/10.1103/PhysRevX.6.041065}{{\em Phys. Rev. X}
  {\bfseries 6} (2016) 041065},
  \href{http://arxiv.org/abs/1605.07331}{{\ttfamily arXiv:1605.07331
  [cond-mat.stat-mech]}}.

\bibitem{Doyon2019a}
B.~Doyon, ``{Lecture notes on Generalised Hydrodynamics},''
  \href{http://dx.doi.org/10.21468/SciPostPhysLectNotes.18}{{\em SciPost Phys.
  Lect. Notes} (2020) 18}, \href{http://arxiv.org/abs/1912.08496}{{\ttfamily
  arXiv:1912.08496}}.

\bibitem{Essler2022}
F.~H. Essler, ``{A short introduction to Generalized Hydrodynamics},''
  \href{http://dx.doi.org/10.1016/j.physa.2022.127572}{{\em Phys. A Stat. Mech.
  its Appl.} (2022) 127572}.

\bibitem{Bastianello2022}
A.~Bastianello, B.~Bertini, B.~Doyon, and R.~Vasseur, ``{Introduction to the
  Special Issue on Emergent Hydrodynamics in Integrable Many-Body Systems},''
  \href{http://dx.doi.org/10.1088/1742-5468/ac3e6a}{{\em J. Stat. Mech. Theory
  Exp.} {\bfseries 2022} (2022) 014001}.

\bibitem{Doyon2017b}
B.~Doyon and H.~Spohn, ``{Drude Weight for the Lieb-Liniger Bose Gas},''
  \href{http://dx.doi.org/10.21468/SciPostPhys.3.6.039}{{\em SciPost Phys.}
  {\bfseries 3} (2017) 039}, \href{http://arxiv.org/abs/1705.08141}{{\ttfamily
  arXiv:1705.08141}}.

\bibitem{Ilievski2017b}
E.~Ilievski and J.~{De Nardis}, ``{Ballistic transport in the one-dimensional
  Hubbard model: The hydrodynamic approach},''
  \href{http://dx.doi.org/10.1103/PhysRevB.96.081118}{{\em Phys. Rev. B}
  {\bfseries 96} (2017) 081118},
  \href{http://arxiv.org/abs/1706.05931}{{\ttfamily arXiv:1706.05931}}.

\bibitem{2018PhRvB..97d5407B}
V.~B. {Bulchandani}, R.~{Vasseur}, C.~{Karrasch}, and J.~E. {Moore},
  ``{Bethe-Boltzmann hydrodynamics and spin transport in the XXZ chain},''
  \href{http://dx.doi.org/10.1103/PhysRevB.97.045407}{{\em Phys. Rev. B}
  {\bfseries 97} (2018) 045407},
  \href{http://arxiv.org/abs/1702.06146}{{\ttfamily arXiv:1702.06146
  [cond-mat.stat-mech]}}.

\bibitem{2019ScPP....6....5U}
A.~{Urichuk}, Y.~{Oez}, A.~{Kl{\"u}mper}, and J.~{Sirker}, ``{The spin Drude
  weight of the XXZ chain and generalized hydrodynamics},''
  \href{http://dx.doi.org/10.21468/SciPostPhys.6.1.005}{{\em SciPost Phys.}
  {\bfseries 6} (2019) 005}.

\bibitem{2022JSMTE2022a4002D}
J.~{De Nardis}, B.~{Doyon}, M.~{Medenjak}, and M.~{Panfil}, ``{Correlation
  functions and transport coefficients in generalised hydrodynamics},''
  \href{http://dx.doi.org/10.1088/1742-5468/ac3658}{{\em J. Stat. Mech. Theor.
  Exp.} {\bfseries 2022} (2022) 014002},
  \href{http://arxiv.org/abs/2104.04462}{{\ttfamily arXiv:2104.04462
  [cond-mat.stat-mech]}}.

\bibitem{2017PhRvB..95f0406K}
C.~{Karrasch}, T.~{Prosen}, and F.~{Heidrich-Meisner}, ``{Proposal for
  measuring the finite-temperature Drude weight of integrable systems},''
  \href{http://dx.doi.org/10.1103/PhysRevB.95.060406}{{\em Phys. Rev. B}
  {\bfseries 95} (2017) 060406},
  \href{http://arxiv.org/abs/1611.04832}{{\ttfamily arXiv:1611.04832
  [cond-mat.str-el]}}.

\bibitem{tsvelik_2003}
A.~M. Tsvelik, \href{http://dx.doi.org/10.1017/CBO9780511615832}{{\em Quantum
  Field Theory in Condensed Matter Physics}}.
\newblock Cambridge University Press, 2003.

\bibitem{Giamarchi:743140}
T.~Giamarchi,
  \href{http://dx.doi.org/10.1093/acprof:oso/9780198525004.001.0001}{{\em
  {Quantum physics in one dimension}}}.
\newblock International series of monographs on physics. Clarendon Press,
  Oxford, 2004.

\bibitem{Controzzi2001}
D.~Controzzi, F.~H.~L. Essler, and A.~M. Tsvelik,
  \href{http://dx.doi.org/10.1007/978-94-010-0838-9_2}{``{Dynamical Properties
  of One Dimensional Mott Insulators},''} in {\em New Theoretical Approaches to
  Strongly Correlated Systems}, A.~M. Tsvelik, ed., p.~25–46.
\newblock Springer Netherlands, Dordrecht, 2001.
\newblock \href{http://arxiv.org/abs/cond-mat/0011439}{{\ttfamily
  arXiv:cond-mat/0011439 [cond-mat.str-el]}}.

\bibitem{2005ffsc.book..684E}
F.~H.~L. {Essler} and R.~M. {Konik},
  \href{http://dx.doi.org/10.1142/9789812775344_0020}{``{Application of Massive
  Integrable Quantum Field Theories to Problems in Condensed Matter
  Physics},''} in {\em From Fields to Strings: Circumnavigating Theoretical
  Physics: Ian Kogan Memorial Collection (in 3 Vols)}, {M. Shifman et al.},
  ed., pp.~684--830.
\newblock {World Scientific}, 2005.
\newblock \href{http://arxiv.org/abs/cond-mat/0412421}{{\ttfamily
  arXiv:cond-mat/0412421 [cond-mat.str-el]}}.

\bibitem{2010PhRvL.105s0403C}
J.~I. {Cirac}, P.~{Maraner}, and J.~K. {Pachos}, ``{Cold Atom Simulation of
  Interacting Relativistic Quantum Field Theories},''
  \href{http://dx.doi.org/10.1103/PhysRevLett.105.190403}{{\em \prl} {\bfseries
  105} (2010) 190403}, \href{http://arxiv.org/abs/1006.2975}{{\ttfamily
  arXiv:1006.2975 [cond-mat.str-el]}}.

\bibitem{2010Natur.466..597H}
E.~{Haller}, R.~{Hart}, M.~J. {Mark}, J.~G. {Danzl}, L.~{Reichs{\"o}llner},
  M.~{Gustavsson}, M.~{Dalmonte}, G.~{Pupillo}, and H.-C. {N{\"a}gerl},
  ``{Pinning quantum phase transition for a Luttinger liquid of strongly
  interacting bosons},'' \href{http://dx.doi.org/10.1038/nature09259}{{\em
  Nature} {\bfseries 466} (2010) 597--600},
  \href{http://arxiv.org/abs/1004.3168}{{\ttfamily arXiv:1004.3168
  [cond-mat.quant-gas]}}.

\bibitem{2017Natur.545..323S}
T.~{Schweigler}, V.~{Kasper}, S.~{Erne}, I.~{Mazets}, B.~{Rauer},
  F.~{Cataldini}, T.~{Langen}, T.~{Gasenzer}, J.~{Berges}, and
  J.~{Schmiedmayer}, ``{Experimental characterization of a quantum many-body
  system via higher-order correlations},''
  \href{http://dx.doi.org/10.1038/nature22310}{{\em Nature} {\bfseries 545}
  (2017) 323--326}, \href{http://arxiv.org/abs/1505.03126}{{\ttfamily
  arXiv:1505.03126 [cond-mat.quant-gas]}}.

\bibitem{2023arXiv230316221W}
E.~{Wybo}, A.~{Bastianello}, M.~{Aidelsburger}, I.~{Bloch}, and M.~{Knap},
  ``{Preparing and Analyzing Solitons in the sine-Gordon Model with Quantum Gas
  Microscopes},'' \href{http://dx.doi.org/10.48550/arXiv.2303.16221}{{\em arXiv
  e-prints} }, \href{http://arxiv.org/abs/2303.16221}{{\ttfamily
  arXiv:2303.16221 [cond-mat.quant-gas]}}.

\bibitem{2021NuPhB.96815445R}
A.~{Roy}, D.~{Schuricht}, J.~{Hauschild}, F.~{Pollmann}, and H.~{Saleur},
  ``{The quantum sine-Gordon model with quantum circuits},''
  \href{http://dx.doi.org/10.1016/j.nuclphysb.2021.115445}{{\em Nucl. Phys. B}
  {\bfseries 968} (2021) 115445},
  \href{http://arxiv.org/abs/2007.06874}{{\ttfamily arXiv:2007.06874
  [quant-ph]}}.

\bibitem{Wybo2022}
E.~Wybo, M.~Knap, and A.~Bastianello, ``{Quantum sine-Gordon dynamics in
  coupled spin chains},''
  \href{http://dx.doi.org/10.1103/PhysRevB.106.075102}{{\em Phys. Rev. B}
  {\bfseries 106} (2022) 075102},
  \href{http://arxiv.org/abs/2203.09530}{{\ttfamily arXiv:2203.09530}}.

\bibitem{1977CMaPh..55..183Z}
A.~B. {Zamolodchikov}, ``{Exact two-particle S-matrix of quantum sine-Gordon
  solitons},'' \href{http://dx.doi.org/10.1007/BF01626520}{{\em Commun. Math.
  Phys.} {\bfseries 55} (1977) 183--186}.

\bibitem{1979AnPhy.120..253Z}
A.~B. {Zamolodchikov} and A.~B. {Zamolodchikov}, ``{Factorized S-matrices in
  two dimensions as the exact solutions of certain relativistic quantum field
  theory models},'' \href{http://dx.doi.org/10.1016/0003-4916(79)90391-9}{{\em
  Ann. Phys.} {\bfseries 120} (1979) 253--291}.

\bibitem{ZAMOLODCHIKOV1991391}
A.~Zamolodchikov, ``{On the thermodynamic Bethe ansatz equations for
  reflectionless ADE scattering theories},''
  \href{http://dx.doi.org/https://doi.org/10.1016/0370-2693(91)91737-G}{{\em
  Phys. Lett. B} {\bfseries 253} (1991) 391--394}.

\bibitem{Tateo:1994pb}
R.~Tateo, ``{The sine-Gordon model as $\frac{\mathcal{S}\mathcal{O}(2n)_1
  \times \mathcal{SO}(2n)_1}{\mathcal{SO}(2n)_2}$ perturbed coset theory and
  generalizations},'' \href{http://dx.doi.org/10.1142/S0217751X95000656}{{\em
  Int. J. Mod. Phys. A} {\bfseries 10} (1995) 1357--1376},
  \href{http://arxiv.org/abs/hep-th/9405197}{{\ttfamily arXiv:hep-th/9405197}}.

\bibitem{Bertini:2019lzy}
B.~Bertini, L.~Piroli, and M.~Kormos, ``{Transport in the sine-Gordon field
  theory: from generalized hydrodynamics to semiclassics},''
  \href{http://dx.doi.org/10.1103/PhysRevB.100.035108}{{\em Phys. Rev. B}
  {\bfseries 100} (2019) 035108},
  \href{http://arxiv.org/abs/1904.02696}{{\ttfamily arXiv:1904.02696
  [cond-mat.stat-mech]}}.

\bibitem{1995PhLB..355..157T}
R.~{Tateo}, ``{New functional dilogarithm identities and sine-Gordon
  Y-systems},'' \href{http://dx.doi.org/10.1016/0370-2693(95)00751-6}{{\em
  Phys. Lett. B} {\bfseries 355} (1995) 157--164},
  \href{http://arxiv.org/abs/hep-th/9505022}{{\ttfamily arXiv:hep-th/9505022
  [hep-th]}}.

\bibitem{yangyang1969}
C.~N. Yang and C.~P. Yang, ``{Thermodynamics of a One‐Dimensional System of
  Bosons with Repulsive Delta‐Function Interaction},''
  \href{http://dx.doi.org/10.1063/1.1664947}{{\em J. Math. Phys.} {\bfseries
  10} (2003) 1115--1122}.

\bibitem{takahashi_1999}
M.~Takahashi, \href{http://dx.doi.org/10.1017/CBO9780511524332}{{\em
  Thermodynamics of One-Dimensional Solvable Models}}.
\newblock Cambridge University Press, 1999.

\bibitem{Zamolodchikov:1989cf}
A.~B. Zamolodchikov, ``{Thermodynamic Bethe Ansatz in Relativistic Models.
  Scaling Three State Potts and Lee-yang Models},''
  \href{http://dx.doi.org/10.1016/0550-3213(90)90333-9}{{\em Nucl. Phys. B}
  {\bfseries 342} (1990) 695--720}.

\bibitem{SM}
Supplemental Material describing (1) details of the TBA system and the dressing
  equations, (2) the bipartitioning protocol, and (3) the analytic evaluation
  of limiting cases.

\bibitem{inpreparation}
{B. C. Nagy, M. Kormos and G. Tak\'acs}, {in preparation}.

\bibitem{boulat2019full}
E.~{Boulat}, ``{Full exact solution of the out-of-equilibrium boundary sine
  Gordon model},'' {\em arXiv e-prints} ,
  \href{http://arxiv.org/abs/1912.03872}{{\ttfamily arXiv:1912.03872
  [cond-mat.str-el]}}.

\bibitem{1991JPhA...24.3111K}
A.~{Kl\"umper}, M.~T. {Batchelor}, and P.~A. {Pearce}, ``{Central charges of
  the 6- and 19-vertex models with twisted boundary conditions},''
  \href{http://dx.doi.org/10.1088/0305-4470/24/13/025}{{\em J. Phys. A Math.
  Gen.} {\bfseries 24} (1991) 3111--3133}.

\bibitem{1995NuPhB.438..413D}
C.~{Destri} and H.~J. {de Vega}, ``{Unified approach to Thermodynamic Bethe
  Ansatz and finite size corrections for lattice models and field theories},''
  \href{http://dx.doi.org/10.1016/0550-3213(94)00547-R}{{\em Nucl. Phys. B}
  {\bfseries 438} (1995) 413--454},
  \href{http://arxiv.org/abs/hep-th/9407117}{{\ttfamily arXiv:hep-th/9407117
  [hep-th]}}.

\bibitem{2020PhRvX..10a1054B}
M.~{Borsi}, B.~{Pozsgay}, and L.~{Pristy{\'a}k}, ``{Current Operators in Bethe
  Ansatz and Generalized Hydrodynamics: An Exact Quantum-Classical
  Correspondence},'' \href{http://dx.doi.org/10.1103/PhysRevX.10.011054}{{\em
  Phys. Rev. X} {\bfseries 10} (2020) 011054},
  \href{http://arxiv.org/abs/1908.07320}{{\ttfamily arXiv:1908.07320
  [cond-mat.stat-mech]}}.

\bibitem{2020PhRvL.125g0602P}
B.~{Pozsgay}, ``{Algebraic Construction of Current Operators in Integrable Spin
  Chains},'' \href{http://dx.doi.org/10.1103/PhysRevLett.125.070602}{{\em Phys.
  Rev. Lett.} {\bfseries 125} (2020) 070602},
  \href{http://arxiv.org/abs/2005.06242}{{\ttfamily arXiv:2005.06242
  [cond-mat.stat-mech]}}.

\bibitem{2019PhRvL.122i0601S}
M.~{Schemmer}, I.~{Bouchoule}, B.~{Doyon}, and J.~{Dubail}, ``{Generalized
  Hydrodynamics on an Atom Chip},''
  \href{http://dx.doi.org/10.1103/PhysRevLett.122.090601}{{\em Phys. Rev.
  Lett.} {\bfseries 122} (2019) 090601},
  \href{http://arxiv.org/abs/1810.07170}{{\ttfamily arXiv:1810.07170
  [cond-mat.quant-gas]}}.

\bibitem{2018PhRvL.120q3601P}
M.~{Pigneur}, T.~{Berrada}, M.~{Bonneau}, T.~{Schumm}, E.~{Demler}, and
  J.~{Schmiedmayer}, ``{Relaxation to a Phase-Locked Equilibrium State in a
  One-Dimensional Bosonic Josephson Junction},''
  \href{http://dx.doi.org/10.1103/PhysRevLett.120.173601}{{\em Phys. Rev.
  Lett.} {\bfseries 120} (2018) 173601},
  \href{http://arxiv.org/abs/1711.06635}{{\ttfamily arXiv:1711.06635
  [quant-ph]}}.

\bibitem{2019OExpr..2733474T}
M.~{Tajik}, B.~{Rauer}, T.~{Schweigler}, F.~{Cataldini}, J.~{Sabino}, F.~S.
  {M{\o}ller}, S.-C. {Ji}, I.~E. {Mazets}, and J.~{Schmiedmayer}, ``{Designing
  arbitrary one-dimensional potentials on an atom chip},''
  \href{http://dx.doi.org/10.1364/OE.27.033474}{{\em Opt. Express} {\bfseries
  27} (2019) 33474}, \href{http://arxiv.org/abs/1908.01563}{{\ttfamily
  arXiv:1908.01563 [cond-mat.quant-gas]}}.

\bibitem{Doyon2019b}
B.~Doyon and J.~Myers, ``{Fluctuations in Ballistic Transport from Euler
  Hydrodynamics},'' \href{http://dx.doi.org/10.1007/s00023-019-00860-w}{{\em
  Ann. Henri Poincar{\'{e}}} {\bfseries 21} (2020) 255--302},
  \href{http://arxiv.org/abs/1902.00320}{{\ttfamily arXiv:1902.00320}}.

\bibitem{Myers2020}
J.~Myers, J.~Bhaseen, R.~J. Harris, and B.~Doyon, ``{Transport fluctuations in
  integrable models out of equilibrium},''
  \href{http://dx.doi.org/10.21468/SciPostPhys.8.1.007}{{\em SciPost Phys.}
  {\bfseries 8} (2020) 007}, \href{http://arxiv.org/abs/1812.02082}{{\ttfamily
  arXiv:1812.02082}}.

\bibitem{DelVecchio2023}
G.~Del Vecchio Del~Vecchio, M.~Kormos, B.~Doyon, and A.~Bastianello, ``{Exact
  large-scale fluctuations of the phase field in the sine-Gordon model},''
  \href{http://arxiv.org/abs/2305.10495}{{\ttfamily arXiv:2305.10495}}.

\bibitem{Bulchandani2021}
V.~B. Bulchandani, S.~Gopalakrishnan, and E.~Ilievski, ``{Superdiffusion in
  spin chains},'' \href{http://dx.doi.org/10.1088/1742-5468/ac12c7}{{\em J.
  Stat. Mech. Theory Exp.} {\bfseries 2021} (2021) 084001},
  \href{http://arxiv.org/abs/2103.01976}{{\ttfamily arXiv:2103.01976}}.

\end{thebibliography}\endgroup


\providecommand{\href}[2]{#2}\begingroup\raggedright\begin{thebibliography}{10}

\bibitem{inpreparation}
{B. C. Nagy, M. Kormos and G. Tak\'acs}, {in preparation}.

\bibitem{ZAMOLODCHIKOV1979253}
A.~B. Zamolodchikov and A.~B. Zamolodchikov, ``{Factorized S-matrices in two
  dimensions as the exact solutions of certain relativistic quantum field
  theory models},''
  \href{http://dx.doi.org/https://doi.org/10.1016/0003-4916(79)90391-9}{{\em
  Ann. Phys.} {\bfseries 120} (1979) 253--291}.

\bibitem{Tateo:1994pb}
R.~Tateo, ``{The sine-Gordon model as $\frac{\mathcal{S}\mathcal{O}(2n)_1
  \times \mathcal{SO}(2n)_1}{\mathcal{SO}(2n)_2}$ perturbed coset theory and
  generalizations},'' \href{http://dx.doi.org/10.1142/S0217751X95000656}{{\em
  Int. J. Mod. Phys. A} {\bfseries 10} (1995) 1357--1376},
  \href{http://arxiv.org/abs/hep-th/9405197}{{\ttfamily arXiv:hep-th/9405197}}.

\bibitem{Feher:2011aa}
G.~Feh\'er and G.~Tak\'acs, ``{Sine-Gordon form factors in finite volume},''
  \href{http://dx.doi.org/10.1016/j.nuclphysb.2011.06.020}{{\em Nucl. Phys. B}
  {\bfseries 852} (2011) 441--467},
  \href{http://arxiv.org/abs/1106.1901}{{\ttfamily arXiv:1106.1901 [hep-th]}}.

\bibitem{2019PhRvB.100c5108B}
B.~{Bertini}, L.~{Piroli}, and M.~{Kormos}, ``{Transport in the sine-Gordon
  field theory: From generalized hydrodynamics to semiclassics},''
  \href{http://dx.doi.org/10.1103/PhysRevB.100.035108}{{\em Phys. Rev. B}
  {\bfseries 100} (2019) 035108},
  \href{http://arxiv.org/abs/1904.02696}{{\ttfamily arXiv:1904.02696
  [cond-mat.stat-mech]}}.

\bibitem{takahashi_1999}
M.~Takahashi, \href{http://dx.doi.org/10.1017/CBO9780511524332}{{\em
  Thermodynamics of One-Dimensional Solvable Models}}.
\newblock Cambridge University Press, 1999.

\bibitem{yangyang1969}
C.~N. Yang and C.~P. Yang, ``{Thermodynamics of a One‐Dimensional System of
  Bosons with Repulsive Delta‐Function Interaction},''
  \href{http://dx.doi.org/10.1063/1.1664947}{{\em J. Math. Phys.} {\bfseries
  10} (2003) 1115--1122}.

\bibitem{Zamolodchikov:1989cf}
A.~B. Zamolodchikov, ``{Thermodynamic Bethe Ansatz in Relativistic Models.
  Scaling Three State Potts and Lee-yang Models},''
  \href{http://dx.doi.org/10.1016/0550-3213(90)90333-9}{{\em Nucl. Phys. B}
  {\bfseries 342} (1990) 695--720}.

\bibitem{1969JMP....10.1115Y}
C.~N. {Yang} and C.~P. {Yang}, ``{Thermodynamics of a One-Dimensional System of
  Bosons with Repulsive Delta-Function Interaction},''
  \href{http://dx.doi.org/10.1063/1.1664947}{{\em J. Math. Phys.} {\bfseries
  10} (1969) 1115--1122}.

\bibitem{ZAMOLODCHIKOV1991391}
A.~Zamolodchikov, ``{On the thermodynamic Bethe ansatz equations for
  reflectionless ADE scattering theories},''
  \href{http://dx.doi.org/https://doi.org/10.1016/0370-2693(91)91737-G}{{\em
  Phys. Lett. B} {\bfseries 253} (1991) 391--394}.

\bibitem{1972PThPh..47...69T}
M.~{Takahashi}, ``{One-Dimensional Hubbard Model at Finite Temperature},''
  \href{http://dx.doi.org/10.1143/PTP.47.69}{{\em Prog. Theor. Phys.}
  {\bfseries 47} (1972) 69--82}.

\bibitem{1995NuPhB.438..413D}
C.~{Destri} and H.~J. {de Vega}, ``{Unified approach to Thermodynamic Bethe
  Ansatz and finite size corrections for lattice models and field theories},''
  \href{http://dx.doi.org/10.1016/0550-3213(94)00547-R}{{\em Nucl. Phys. B}
  {\bfseries 438} (1995) 413--454},
  \href{http://arxiv.org/abs/hep-th/9407117}{{\ttfamily arXiv:hep-th/9407117
  [hep-th]}}.

\bibitem{Doyon_2020}
B.~Doyon, ``{Lecture notes on Generalised Hydrodynamics},'' {\em SciPost Phys.
  Lect. Notes} (2020) 18, \href{http://arxiv.org/abs/1912.08496}{{\ttfamily
  arXiv:1912.08496}}.

\bibitem{Mossel_2012}
J.~Mossel and J.-S. Caux, ``{Generalized TBA and generalized Gibbs},''
  \href{http://dx.doi.org/10.1088/1751-8113/45/25/255001}{{\em J. Phys. A Math.
  Theor.} {\bfseries 45} no.~25, (May, 2012) 255001}.

\bibitem{2016PhRvL.117t7201B}
B.~{Bertini}, M.~{Collura}, J.~{De Nardis}, and M.~{Fagotti}, ``{Transport in
  Out-of-Equilibrium XXZ Chains: Exact Profiles of Charges and Currents},''
  \href{http://dx.doi.org/10.1103/PhysRevLett.117.207201}{{\em Phys. Rev.
  Lett.} {\bfseries 117} (2016) 207201},
  \href{http://arxiv.org/abs/1605.09790}{{\ttfamily arXiv:1605.09790
  [cond-mat.stat-mech]}}.

\bibitem{2016PhRvX...6d1065C}
O.~A. {Castro-Alvaredo}, B.~{Doyon}, and T.~{Yoshimura}, ``{Emergent
  Hydrodynamics in Integrable Quantum Systems Out of Equilibrium},''
  \href{http://dx.doi.org/10.1103/PhysRevX.6.041065}{{\em Phys. Rev. X}
  {\bfseries 6} (2016) 041065},
  \href{http://arxiv.org/abs/1605.07331}{{\ttfamily arXiv:1605.07331
  [cond-mat.stat-mech]}}.

\bibitem{Piroli_2017}
L.~{Piroli}, J.~{De Nardis}, M.~{Collura}, B.~{Bertini}, and M.~{Fagotti},
  ``{Transport in out-of-equilibrium XXZ chains: Nonballistic behavior and
  correlation functions},''
  \href{http://dx.doi.org/10.1103/PhysRevB.96.115124}{{\em Phys. Rev. B}
  {\bfseries 96} (2017) 115124},
  \href{http://arxiv.org/abs/1706.00413}{{\ttfamily arXiv:1706.00413
  [cond-mat.stat-mech]}}.

\bibitem{Doyon_2017}
B.~Doyon and H.~Spohn, ``{Drude Weight for the Lieb-Liniger Bose Gas},''
  \href{http://dx.doi.org/10.21468/scipostphys.3.6.039}{{\em {SciPost} Physics}
  {\bfseries 3} (2017) }.

\bibitem{2019ScPP....6....5U}
A.~{Urichuk}, Y.~{Oez}, A.~{Kl{\"u}mper}, and J.~{Sirker}, ``{The spin Drude
  weight of the XXZ chain and generalized hydrodynamics},''
  \href{http://dx.doi.org/10.21468/SciPostPhys.6.1.005}{{\em SciPost Phys.}
  {\bfseries 6} (2019) 005}.

\bibitem{2022JSMTE2022a4002D}
J.~{De Nardis}, B.~{Doyon}, M.~{Medenjak}, and M.~{Panfil}, ``{Correlation
  functions and transport coefficients in generalised hydrodynamics},''
  \href{http://dx.doi.org/10.1088/1742-5468/ac3658}{{\em J. Stat. Mech. Theor.
  Exp.} {\bfseries 2022} (2022) 014002},
  \href{http://arxiv.org/abs/2104.04462}{{\ttfamily arXiv:2104.04462
  [cond-mat.stat-mech]}}.

\bibitem{Bertini:2019lzy}
B.~Bertini, L.~Piroli, and M.~Kormos, ``{Transport in the sine-Gordon field
  theory: from generalized hydrodynamics to semiclassics},''
  \href{http://dx.doi.org/10.1103/PhysRevB.100.035108}{{\em Phys. Rev. B}
  {\bfseries 100} (2019) 035108},
  \href{http://arxiv.org/abs/1904.02696}{{\ttfamily arXiv:1904.02696
  [cond-mat.stat-mech]}}.

\bibitem{Andrews:1984af}
G.~E. Andrews, R.~J. Baxter, and P.~J. Forrester, ``{Eight vertex SOS model and
  generalized Rogers-Ramanujan type identities},''
  \href{http://dx.doi.org/10.1007/BF01014383}{{\em J. Statist. Phys.}
  {\bfseries 35} (1984) 193--266}.

\bibitem{Klassen:1990dx}
T.~R. Klassen and E.~Melzer, ``{The Thermodynamics of purely elastic scattering
  theories and conformal perturbation theory},''
  \href{http://dx.doi.org/10.1016/0550-3213(91)90159-U}{{\em Nucl. Phys. B}
  {\bfseries 350} (1991) 635--689}.

\bibitem{1995PhLB..355..157T}
R.~{Tateo}, ``{New functional dilogarithm identities and sine-Gordon
  Y-systems},'' \href{http://dx.doi.org/10.1016/0370-2693(95)00751-6}{{\em
  Phys. Lett. B} {\bfseries 355} (1995) 157--164},
  \href{http://arxiv.org/abs/hep-th/9505022}{{\ttfamily arXiv:hep-th/9505022
  [hep-th]}}.

\bibitem{2012arXiv1212.6853N}
T.~{Nakanishi} and S.~{Stella}, ``{Wonder of sine-Gordon Y-systems},''
  \href{http://dx.doi.org/10.1090/tran/6505}{{\em {Trans. Amer. Math. Soc.}}
  {\bfseries 368} (2016) 6835--6886},
  \href{http://arxiv.org/abs/1212.6853}{{\ttfamily arXiv:1212.6853 [math.QA]}}.

\end{thebibliography}\endgroup
\end{document}


\newcommand{\BN}[1]{\textcolor{red}{[BN: #1]}}
\newcommand{\KM}[1]{\textcolor{orange}{[KM: #1]}}
\newcommand{\GT}[1]{\textcolor{violet}{[GT: #1]}}

\newcommand{\MK}[1]{\textcolor{orange}{#1}}

\newcommand{\op}[1]{\mathcal{#1}}
\newcommand{\de}{\mathrm d}
\newcommand{\Tr}{\operatorname{Tr}}
\newcommand{\ket}[1]{\left |#1 \right \rangle }
\renewcommand\th          {\theta}

\def\bes{\begin{subequations}}
\def\esu{\end{subequations}}
\newcommand\be{\begin{equation}}
\newcommand\ee{\end{equation}}

\title{Supplemental Material for ``Thermodynamics and fractal Drude weights in the sine-Gordon model''}

\author{Botond C. Nagy}
\affiliation{Department of Theoretical Physics, Institute of Physics, Budapest University of Technology and Economics, H-1111 Budapest, M{\H u}egyetem rkp. 3.}
\affiliation{
BME-MTA Momentum Statistical Field Theory Research Group, Institute of Physics, Budapest University of Technology and Economics, H-1111 Budapest, M{\H u}egyetem rkp. 3.}
\author{M\'arton Kormos}
\affiliation{Department of Theoretical Physics, Institute of Physics, Budapest University of Technology and Economics, H-1111 Budapest, M{\H u}egyetem rkp. 3.}
\affiliation{MTA-BME Quantum Correlations Group (ELKH), Institute of Physics, Budapest University of Technology and Economics, H-1111 Budapest, M{\H u}egyetem rkp. 3.}
\affiliation{
BME-MTA Momentum Statistical Field Theory Research Group, Institute of Physics, Budapest University of Technology and Economics, H-1111 Budapest, M{\H u}egyetem rkp. 3.}
\author{G\'abor Tak\'acs}
\affiliation{Department of Theoretical Physics, Institute of Physics, Budapest University of Technology and Economics, H-1111 Budapest, M{\H u}egyetem rkp. 3.}
\affiliation{MTA-BME Quantum Correlations Group (ELKH), Institute of Physics, Budapest University of Technology and Economics, H-1111 Budapest, M{\H u}egyetem rkp. 3.}
\affiliation{
BME-MTA Momentum Statistical Field Theory Research Group, Institute of Physics, Budapest University of Technology and Economics, H-1111 Budapest, M{\H u}egyetem rkp. 3.}

\date{24th May, 2023}
\maketitle
\tableofcontents

\section{Thermodynamic Bethe Ansatz in sine-Gordon model}

In this section, we summarise the TBA equations of the sine-Gordon model with a sketch of their derivation (for more details c.f.~\cite{inpreparation}). The sine-Gordon spectrum consists of a kink/antikink pair with mass $M$ and $n_B=\lfloor1/\xi\rfloor$ breathers with masses $M_{B_i}$, $i=1,\dots n_B$\footnote{When $1/\xi$ is integer, $n_B=1/\xi-1$.}. The energy and momentum of these particles can be given in terms of their masses $M_a$ as $e_a=M_a\cosh\theta$ and $p_a=M_a\sinh\theta$ where $\theta$ is the so-called rapidity variable. 

\subsection{Scattering amplitudes and finite volume spectrum}
Soliton scattering is described by the following two-particle amplitudes \cite{ZAMOLODCHIKOV1979253}:
\begin{align}
&S_{++}^{++}(\theta)=S_{--}^{--}(\theta)=S_0(\theta)\,,\quad\quad
S_{+-}^{+-}(\theta)=S_{-+}^{-+}(\theta)=S_T(\theta)S_0(\theta)\,,\quad\quad
S_{+-}^{-+}(\theta)=S_{-+}^{+-}(\theta)=S_R(\theta)S_0(\theta)\,,\nonumber\\
&S_T(\theta) = \frac{\sinh\left(\frac{\theta}{\xi}\right)}{\sinh\left(\frac{i\pi-\theta}{\xi}\right)}\,,\quad\quad 
S_R(\theta) = \frac{i\sin\left(\frac{\pi}{\xi}\right)}{\sinh\left(\frac{i\pi-\theta}{\xi}\right)}\,,\quad\quad
S_0(\theta) = -\exp\left(\int\limits_{-\infty}^{\infty} \frac{\mathrm{d}t}{t}
                  \frac{\sinh\left(\frac{t\pi}{2}(\xi-1)\right)}{2\sinh\left(\frac{\pi\xi t}{2}\right)\cosh\left(\frac{\pi t}{2}\right)}{e}^{i\theta t}\right)\,,
\label{SS_scattering_matrix}
\end{align}
where $+/-$ stands for kinks/antikinks, with $\theta$ denoting the difference of their rapidities. The scattering amplitude between a kink/antikink and a breather is 
\begin{equation}
S_{\pm,B_n}(\theta) = \frac{\sinh(\theta)+i \cos\left(\frac{n\pi\xi}{2}\right)}{\sinh(\theta)-i \cos\left(\frac{n\pi\xi}{2}\right)}
                \prod_{k=1}^{n-1}
                      \frac{\sin^2\left(\frac{(n-2k)\pi\xi}{4}-\frac{\pi}{4}+i\frac{\theta}{2}\right)}{\sin^2\left(\frac{(n-2k)\pi\xi}{4}-\frac{\pi}{4}-i\frac{\theta}{2}\right)}\,,
   \label{sB_scattering_matrix}
\end{equation}
while between two breathers 
\begin{equation}
\begin{split}
    S_{B_n,B_m}(\theta) &= \frac{\sinh(\theta)+i \sin\left(\frac{(n+m)\pi\xi}{2}\right)}{\sinh(\theta)-i \sin\left(\frac{(n+m)\pi\xi}{2}\right)}
                         \frac{\sinh(\theta)+i \sin\left(\frac{(n-m)\pi\xi}{2}\right)}{\sinh(\theta)-i \sin\left(\frac{(n-m)\pi\xi}{2}\right)} \times \\
                 &\times \prod_{k=1}^{\text{min}(n,m)-1}
                         \frac{\sin^2\left(\frac{(\left|m-n\right|-2k)\pi\xi}{4}+i\frac{\theta}{2}\right)}{\sin^2\left(\frac{(\left|m-n\right|-2k)\pi\xi}{4}-i\frac{\theta}{2}\right)}
                         \frac{\cos^2\left(\frac{(m+n-2k)\pi\xi}{4}+i\frac{\theta}{2}\right)}{\cos^2\left(\frac{(m+n-2k)\pi\xi}{4}-i\frac{\theta}{2}\right)}\,.
    \label{SM:S_BB}
\end{split}
\end{equation}
Note that for integer values of $1/\xi$ the kink-antikink reflection amplitude $S_R$ vanishes, corresponding to reflectionless (purely transmissive) scattering.

Up to exponentially small corrections, the spectrum in a finite volume $L$ with periodic boundary conditions can be computed from the following quantisation conditions \cite{Tateo:1994pb, Feher:2011aa, 2019PhRvB.100c5108B}:
\begin{align}
    {e}^{i M_{B_i} L \sinh\theta_{B_i}^{(j)}}  
    \underset{(k,l)\neq (i,j)}{\prod_{k=1}^{n_B}\prod_{l=1}^{N_{B_k}}}
    S_{B_i,B_k}\left(\theta_{B_i}^{(j)}-\theta_{B_k}^{(l)}\right) \prod_{k=1}^{N_s} S_{\pm,B_i}\left(\theta_{B_i}^{(j)}-\theta_k\right) &=\phantom{-}1\,,\qquad
    j=1,...,N_{B_i},\quad i=1,...,n_B\,,\nonumber\\
    {e}^{i M L \sinh\theta_k} \Lambda(\theta_k | \{\mu_m\} , \{\theta_l\}) \prod_{i=1}^{n_B} \prod_{j=1}^{N_{B_i}} S_{\pm,B_i}\left(\theta_k-\theta_{B_i}^{(j)}\right) &= -1\,,\qquad
    k=1,...,N_s\,,
    \label{eq:BYquantisation}
\end{align}
written for a state containing $N_s$ kinks/antikinks with rapidities $\theta_l$, $l=1,\dots N_s$ and $N_{B_i}$ breathers of type $B_i$ with rapidities $\theta_{B_i}^{(j)}$, $j=1,\dots,N_{B_i}$. The function 
\begin{equation}
    \Lambda\left(\theta_k | \{\mu_m\}, \{\theta_l\}\right) = \prod_{m=1}^{N_m} 
    \frac{1}{S_T(\mu_m-\theta_k)} 
    \prod_{l=1}^{N_s} S_0(\theta_k-\theta_l)
    \label{eq:Lambda}
\end{equation}
is expressed in terms of auxiliary variables $\mu_r$, $r=1,\dots,N_m$ which we call elementary magnons, that satisfy the relations \cite{Feher:2011aa}
\begin{align}
    \prod_{l=1}^{N_s} \frac{1}{S_T(\mu_r-\theta_l)} = 
    \prod_{\substack{q=1\\q\neq r}}^{N_m} 
    \frac{S_T(\mu_r-\mu_q)}{S_T(\mu_q-\mu_r)}\,,\qquad r=1,\dots,N_m\,.
    \label{eq:bymagnon}
\end{align}
Magnons are auxiliary massless excitations accounting for the kinks' charge degrees of freedom; the state's total topological charge is given by $Q=N_s-2 N_m$. 

\subsection{Thermodynamic limit}
According to the string conjecture, in the thermodynamic limit the elementary magnons can be assumed to arrange into particular patterns called strings, in which the rapidities have the same real part and are equally spaced in the imaginary direction. Eqs. \eqref{eq:bymagnon} are identical to the Bethe Ansatz equations of the XXZ spin chain in its gapless regime, and consequently the string classification is identical to the one for the XXZ chain which relies on the representation of the coupling as a continued fraction
\begin{equation}
    \xi = \frac{1}{n_B+\displaystyle\frac{1}{\nu_1+\displaystyle\frac{1}{\nu_2+...}}} = \frac{1}{n_B+\displaystyle\frac{1}{\alpha}}\,,
    \label{SM:cont_fraction}
\end{equation}
where the $\nu_i$ are natural numbers. The details are rather involved, so we omit them here and refer the interested reader to \cite{takahashi_1999}. It is only important to keep in mind that the magnon strings can be classified into levels with exactly $\nu_l$ magnon strings at level $l$. The end result is that the finite volume spectrum can be parameterised by the rapidities of solitons, breathers and magnon strings. From now, when referring to magnons we mean the magnonic string configurations. 

Following the usual line of thought \cite{yangyang1969,Zamolodchikov:1989cf,takahashi_1999}, in the thermodynamic limit the rapidities can be described by root densities $\rho^{r}_a(\theta)$, where 
\begin{equation}
    L\rho^{\text{r}}_a(\theta)d\theta    
\end{equation}
gives the number of excitations of type $a$ in the interval $[\theta,\theta+d\theta]$, with $a$ going over the breathers, the soliton and the antisoliton for reflectionless points, while in the general case it goes over the breathers, one solitonic excitation and the magnons. One similarly introduces the densities $\rho^{\text{h}}_a(\theta)$ of holes of type $a$ and the total densities of states $\rho^{\text{tot}}_a(\theta)=\rho^{\text{r}}_a(\theta)+\rho^{\text{h}}_a(\theta)$. The quantisation conditions (\ref{eq:BYquantisation},\ref{eq:bymagnon}) together with \eqref{eq:Lambda} then imply the following integral equations 
\begin{equation}
    \rho_a^{\text{tot}} = \rho_a^{\text{r}}+\rho_a^{\text{h}} = \eta_a s_a + \sum_b \eta_a \Phi_{ab}*\rho_b^{\text{r}} \,,
    \label{SM:density_eqs}
\end{equation}
where the source terms are 
\begin{equation}
    s_a=\frac{M_a}{2\pi}\cosh\theta
\label{SM:sources}
\end{equation} 
are understood to be zero for the magnons due their masslessness. The $\Phi_{ab}$ are kernels given by the logarithmic derivatives of the phase shifts, and $\eta_a=\pm 1$ are signs required to make the total densities positive and are nontrivial only for magnonic excitations and are specified later below. The convolution operation is defined as 
\begin{equation}
    (f*g)(\theta) = \int \frac{\mathrm{d}\theta}{2\pi} f(\theta-\theta') g(\theta')\,.
\end{equation}
Using the following conventions for Fourier transformation
\begin{equation}
   f(\theta) = \int\limits_{-\infty}^{\infty} \mathrm{d}t \tilde{f}(t) {e}^{i\theta t}\,, \hspace{1cm} 
   \tilde{f}(t) = \int\limits_{-\infty}^{\infty} \frac{\mathrm{d}\theta}{2\pi} f(\theta) {e}^{-i\theta t}\,,
\end{equation}
the kernels are expressed in Fourier space as
\begin{align}
    \tilde{\Phi}_0(t) &= -\frac{\sinh(\frac{\pi}{2}(1-\xi)t)}{2\sinh(\frac{\pi}{2}\xi t)\cosh(\frac{\pi}{2}t)} \nonumber\\
    \tilde{\Phi}_{A,B_k}(t) &= - \frac{\cosh(\frac{\pi}{2}\xi t)}{\sinh(\frac{\pi}{2}\xi t)} \frac{1}{\cosh(\frac{\pi}{2} t)} \sinh\left(k\frac{\pi}{2}\xi t\right) \nonumber\\
    \tilde{\Phi}_{B_j,B_k}(t) &= \delta_{j,k} - 2 \frac{1}{\cosh(\frac{\pi}{2}t)}\frac{\cosh(\frac{\pi}{2}\xi t)}{\sinh(\frac{\pi}{2}\xi t)}
                          \cosh\left(\frac{\pi}{2}\left( 1- j \xi \right) t\right) \sinh\left(\frac{\pi}{2} k \xi t\right)\,,\, j\geq k\nonumber\\
    \tilde{\Phi}_{A,m_i}(t) &= a\left( \frac{t}{\frac{2}{\pi}(n_B\alpha + 1)}, \ell_{m_i}, v_{m_i} \right)\,,\nonumber\\
    \tilde{\Phi}_{m_i,m_j}(t) &= -\bigg[
        a\left( \frac{t}{\frac{2}{\pi}(n_B\alpha + 1)}, \left|\ell_{m_i}-\ell_{m_p}\right|, v_{m_i}v_{m_j} \right)\nonumber
        +\sum_{k=1}^{\textrm{min}(\ell_{m_i},\ell_{m_p})-1} 2a\left( \frac{t}{\frac{2}{\pi}(n_B\alpha + 1)}, \left|\ell_{m_i}-\ell_{m_p}\right|+2k, v_{m_i}v_{m_j} \right) \nonumber\\
        &+ a\left( \frac{t}{\frac{2}{\pi}(n_B\alpha + 1)}, \ell_{m_i}+\ell_{m_p}, v_{m_i}v_{m_j} \right)\Bigg]\,,
\end{align}
where $\ell_{m_i}$ is the length of the magnonic string species $m_i$, while $v_{m_i}$ is its parity \cite{takahashi_1999}, and
\begin{align}
    a(t,k,+1) &= 
    \begin{cases}
        0 &\textrm{for } k=0\, \textrm{mod}(0,\alpha)\,,\\
       \displaystyle \frac{\sinh[(\hat{k}-\alpha) t]}{\sinh\alpha t}\,,\hspace{0.5cm} \hat{k}=k\ \textrm{mod}\ (0,2\alpha)\ &\textrm{otherwise}\,,
    \end{cases}
    \nonumber\\
    a(t,k,-1) &= 
    \begin{cases}
        0 &\textrm{for } k=0\, \textrm{mod}(0,\alpha)\,,\\
       \displaystyle \frac{\sinh \hat{k} t}{\sinh\alpha t}\,, \hspace{1.5cm} \hat{k}=k\ \textrm{mod}\ (-\alpha,\alpha)\  &\textrm{otherwise}\,.
    \end{cases}
\end{align}

Eq.~(\ref{SM:density_eqs}) still contains two independent functions, the root and the hole densities, as the equation holds for any state containing an extensive number of one-particle states. In the thermodynamic limit, the most probable state of the system is found by minimising the free energy density $f=e-Ts-\mu q$, where $s$ is the Yang-Yang entropy \cite{1969JMP....10.1115Y}, and $\mu$ is the chemical potential conjugate to the topological charge. This procedure leads to the system of non-linear integral equations
\begin{equation}
    \epsilon_a = w_a - \sum_{b}\eta_b \Phi_{ab}*\log(1+e^{-\epsilon_b})
    \label{SM:TBA_coupled_eq}
\end{equation}
in terms of the pseudo-energies 
\begin{equation}
    \epsilon_a = \log\left(\frac{\rho^{\text{h}}_a}{\rho^{\text{r}}_a}\right)
\end{equation}
with the driving terms 
\begin{equation}
    w_a=\frac{M_a}{T} \cosh\theta - \frac{\mu}{T}q_a\,,
\end{equation}
while the topological charge takes the value $q_{\pm}=\pm 1$ for the soliton and the antisoliton, and $q_{m_i}=-2\ell_{m_i}$ for the magnons. It is very important to note that the density relations \eqref{SM:density_eqs} can be obtained by taking the derivative of the TBA equations \eqref{SM:TBA_coupled_eq} with respect to $R=1/T$ with the identification
\begin{equation}
    \rho_a^\mathrm{tot}=\eta_a\frac{1}{2\pi}\frac{\partial\epsilon_a}{\partial R}\,,
\label{eq:density_as_derivative}
\end{equation}
where the signs $\eta_a=\pm 1$ ensure the positivity of the densities.

\subsection{Partially decoupled TBA system}
The TBA equations \eqref{SM:TBA_coupled_eq} couple all the densities of all the excitations. However, they can be partially decoupled, which is especially useful from the point of view of numerical calculations. The decoupling follows the ideas of \cite{ZAMOLODCHIKOV1991391,takahashi_1999}; for the detailed calculation see \cite{inpreparation}. This procedure results in the partially decoupled form of the TBA system 
\begin{equation}
    \epsilon_a = w_a + \sum_b K_{ab} * \left( \sigma_b^{(1)}\epsilon_b - \sigma_b^{(2)} w_b + L_b \right)\, ,
    \label{SM:TBA_eq_decoupled}
\end{equation}
where we defined $L_b(\theta)=\log(1+e^{-\epsilon_b(\theta)})$. It is important to stress that besides the kernels, the magnonic driving terms are also modified by the decoupling procedure and the tables below contain these "decoupled" driving terms. Analogously, the densities satisfy the partially decoupled system
\begin{equation}
    \eta_a\,\rho^{\textrm{tot}}_a =  \frac{\partial_\theta p_a}{2\pi} + \sum_b K_{ab}*\left[\left(\sigma_b^{(1)}\!\!-\vartheta_b\right)\eta_b\,\rho^{\textrm{tot}}_b-\sigma_b^{(2)} \frac{\partial_{\theta}p_b}{2\pi} \right]
    \label{SM:density_eq}
\end{equation}
with $p_a=M_a\sinh\theta$  and the filling fractions $\vartheta_a$ are defined as
\begin{equation}
    \vartheta_a(\theta)=\frac{\rho^\mathrm{r}_a(\theta)}{\rho^\mathrm{tot}_a(\theta)}
    =\frac{1}{1+e^{\epsilon_a}}\,.
\end{equation}
The kernels $K_{ab}$ are specified below, while the driving terms $w_a$, and the factors $\sigma_a^{(1)}$, $\sigma_a^{(2)}$ and $\eta_a$ are given in tables in the next subsection for different cases of the continued fraction decomposition of $\xi$ (\ref{SM:cont_fraction}) with at most two magnonic levels.

\subsection{Kernels and other ingredients of the partially decoupled TBA system}

The diagrams encoding the partially decoupled TBA system \eqref{SM:TBA_eq_decoupled} can be built up from the following six types of building blocks
\begin{figure}[H]
\captionsetup{justification=centering}
    \centering
    \includegraphics[width=0.9\textwidth]{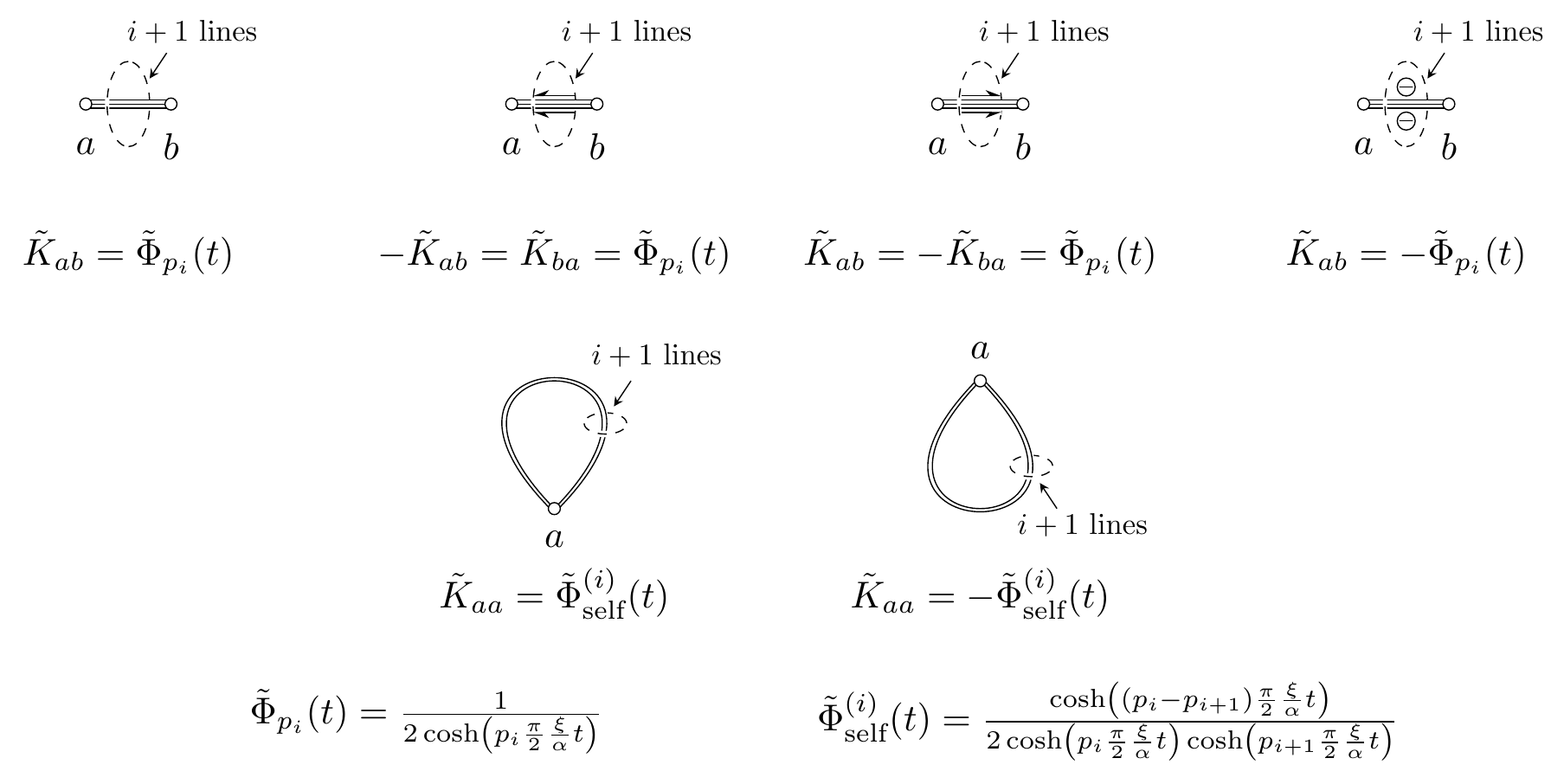}
\end{figure}
The parameters $p_i$ appearing in expressions of the kernels are (for at most two magnonic levels)
\begin{equation}
    p_0=\alpha\,,\ p_1=1\,,\ p_2=1/\nu_2\,.
\end{equation}

\paragraph{Reflectionless points.} For the reflectionless points 
\begin{equation}
    \xi=\displaystyle\frac{1}{n_B-1}\,,
\end{equation} 
the diagram describing the TBA kernels is drawn below, with the associated table giving the driving terms, topological charges and sign factors entering the TBA system.

\begin{figure}[H]
\captionsetup{justification=centering}
\begin{minipage}{0.4\textwidth}
\centering
\includegraphics[height=2.75cm]{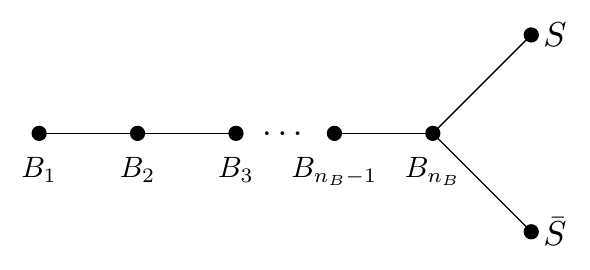}
\end{minipage}
\begin{minipage}{0.55\textwidth}
\centering
\begin{table}[H]
\centering
\begin{tabular}{|l|l|c|c|c|c|c|}
\hline
Excitations & Labels               & $w$                                       & $q$  & $\eta$ & $\sigma^{(1)}$ & $\sigma^{(2)}$ \\ \hline\hline
Breathers   & $B_i$, $i=1,...,n_B$ & ${M_{B_i}} \cosh(\theta)/T$           & $0$  & $+1$   & $+1$           & $+1$           \\ \hline
Soliton     & $S$                  & ${M} \cosh(\theta)/T - {\mu}/{T}$ & $+1$ & $+1$   & $+1$           & $+1$           \\ \hline
Antisoliton & $\bar{S}$            & ${M} \cosh(\theta)/T + {\mu}/{T}$ & $-1$ & $+1$   & $+1$           & $+1$           \\ \hline
\end{tabular}
\end{table}
\end{minipage}
\end{figure}

\paragraph{One magnon level.} For the case of one magnon level corresponding to a coupling with continued fraction expansion
\begin{equation}
    \xi=\displaystyle\frac{1}{n_B+\displaystyle\frac{1}{\nu_1}}\,,
\end{equation}
the TBA system is specified by the diagrams
\begin{figure}[H]
    \centering
    \captionsetup{justification=centering}
    \begin{subfigure}{0.24\textwidth}
    \centering
        \includegraphics[height=3.25cm]{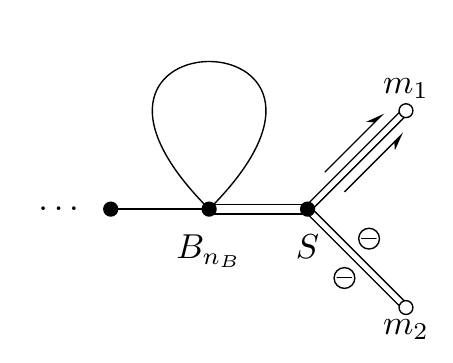}
        \caption{$\nu_1=2$.}
    \end{subfigure}
    \hfill
    \begin{subfigure}{0.29\textwidth}
    \centering
        \includegraphics[height=3.25cm]{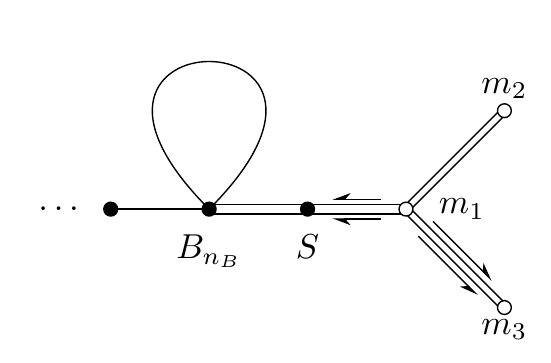}
        \caption{$\nu_1=3$.}
    \end{subfigure}
    \hfill
    \begin{subfigure}{0.45\textwidth}
    \centering
        \includegraphics[height=3.25cm]{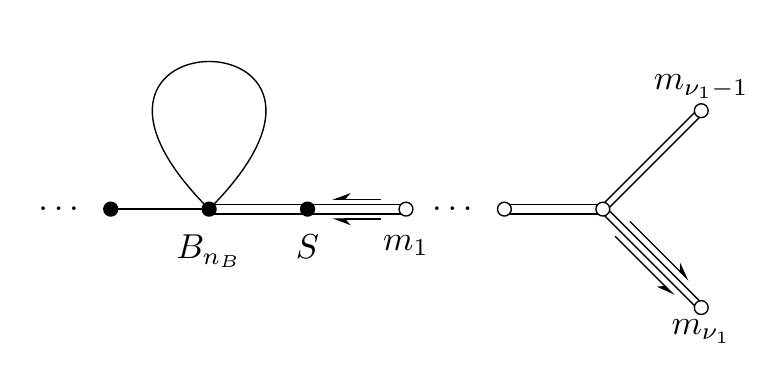}
        \caption{$\nu_1>3$.}
    \end{subfigure}
\end{figure}
together with the following driving terms, signs, and topological charges:
\begin{table}[H]
\centering
\captionsetup{justification=centering}
\begin{tabular}{|l|l|c|c|c|c|c|c|}
\hline
Excitations                      & Labels                       & $w$                             & $q$         & $\eta$ & $v$  & $\sigma^{(1)}$ & $\sigma^{(2)}$ \\ \hline\hline
Breathers                        & $B_i$, $i=1,...,n_B$         & ${M_{B_i}} \cosh(\theta)/T$ & $0$         & $+1$   & $0$  & $+1$           & $+1$           \\ \hline
Soliton                          & $S$                          & ${M} \cosh(\theta)/T$       & $+1$        & $+1$   & $0$  & $0$            & $0$            \\ \hline
First level intermediate magnons & $m_j$, $i=j,...\nu_1-2$      & $0$                             & $-2\cdot j$ & $-1$   & $+1$ & $+1$           & $0$            \\ \hline
First level next-to-last magnon  & $m_{\nu_1-1}$, ($j=\nu_1-1$) & $\nu_1\cdot \mu/T$      & $-2\cdot j$ & $-1$   & $+1$ & $+1$           & $0$            \\ \hline
First level last magnon         & $m_{\nu_1}$                  & $\nu_1\cdot \mu/T$      & $-2$        & $+1$   & $-1$ & $0$            & $0$            \\ \hline
\end{tabular}
\end{table}

\paragraph{Two magnon levels.} For the case of two magnon levels corresponding to a coupling with the continued fraction expansion
\begin{equation}
\xi=\displaystyle\frac{1}{n_B+\displaystyle\frac{1}{\nu_1+\displaystyle\frac{1}{\nu_2}}}\,,    
\end{equation}
the TBA system is specified by the diagrams
\begin{figure}[H]
    \centering
    \captionsetup{justification=centering}
    \begin{subfigure}{0.45\textwidth}
    \centering
        \includegraphics[height=3.5cm]{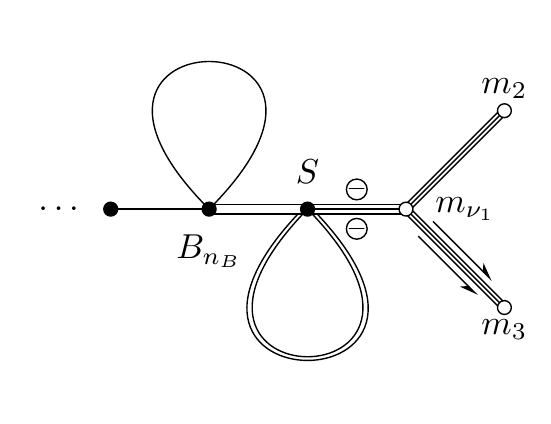}
        \caption{$\nu_1=1$, $\nu_2=2$.}
    \end{subfigure}
    \hfill
    \begin{subfigure}{0.54\textwidth}
    \centering
        \includegraphics[height=3.5cm]{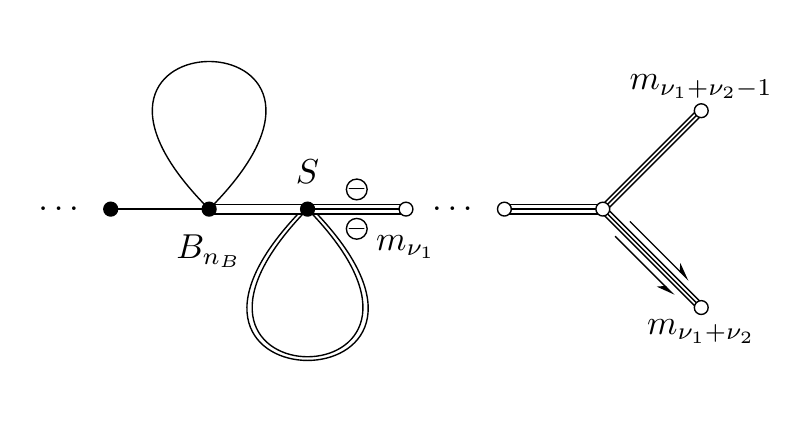}
        \caption{$\nu_1=1$, $\nu_2>2$.}
    \end{subfigure}
\end{figure}
\begin{figure}[H]
    \centering
    \captionsetup{justification=centering}
    \begin{subfigure}{0.45\textwidth}
    \centering
        \includegraphics[height=3.25cm]{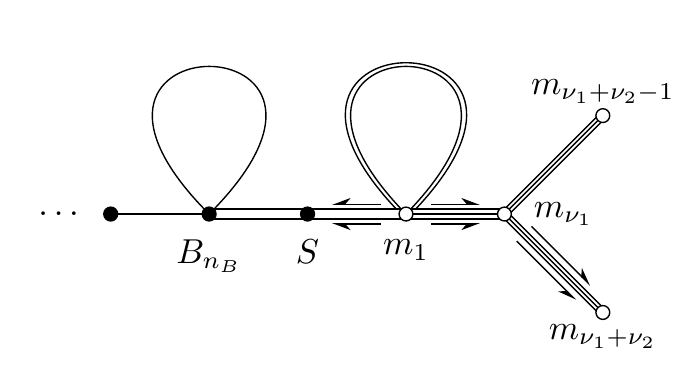}
        \caption{$\nu_1=2$, $\nu_2=2$.}
    \end{subfigure}
    \hfill
    \begin{subfigure}{0.54\textwidth}
    \centering
        \includegraphics[height=3.25cm]{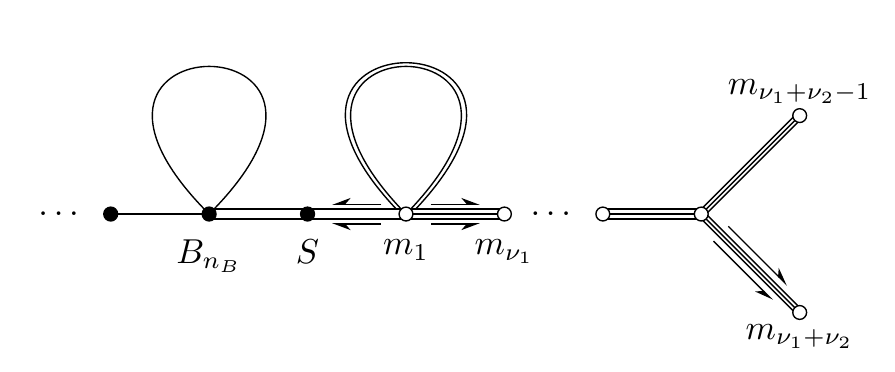}
        \caption{$\nu_1=2$, $\nu_2>2$.}
    \end{subfigure}
\end{figure}
\begin{figure}[H]
    \centering
    \captionsetup{justification=centering}
    \begin{subfigure}{0.45\textwidth}
    \centering
        \includegraphics[height=3.25cm]{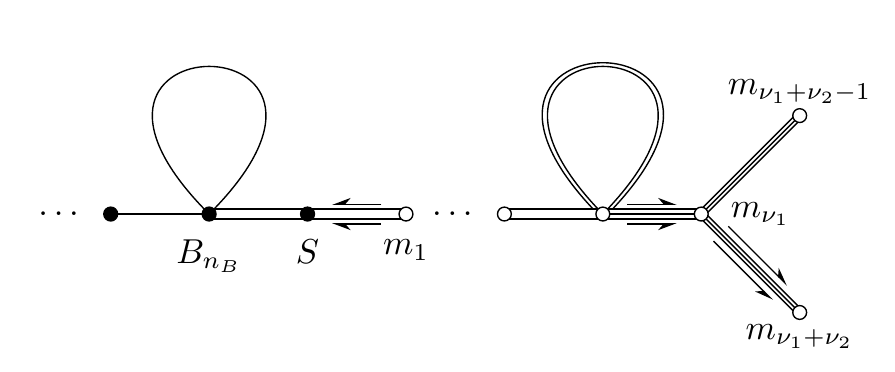}
        \caption{$\nu_1>2$, $\nu_2=2$.}
    \end{subfigure}
    \hfill
    \begin{subfigure}{0.54\textwidth}
    \centering
        \includegraphics[height=3.25cm]{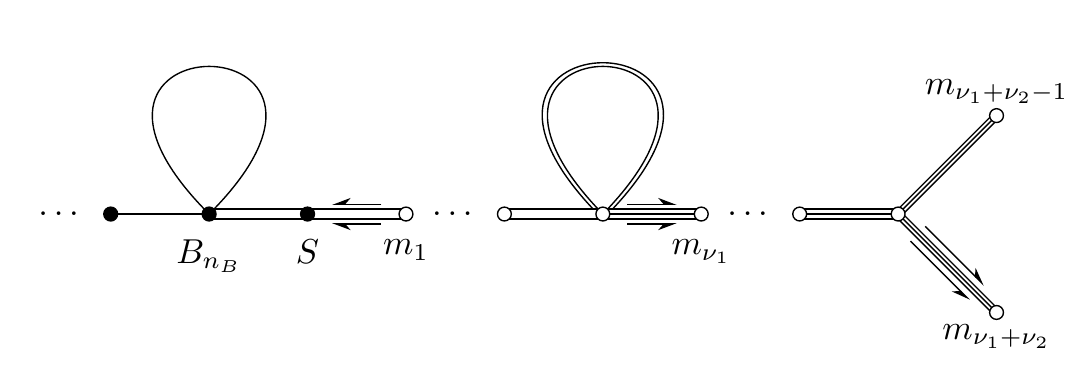}
        \caption{$\nu_1>2$, $\nu_2>2$.}
    \end{subfigure}
\end{figure}
together with the following driving terms, parities, signs and topological charges:
\begin{table}[H]
\centering
\captionsetup{justification=centering}
\begin{tabular}{|l|l|c|c|c|c|c|c|}
\hline
Excitations                       & Labels                           & $w$                                          & $q$                        & $\eta$ & $v$                                                                         & $\sigma^{(1)}$ & $\sigma^{(2)}$ \\ \hline\hline
Breathers                         & $B_i$, $i=1,...,n_B$             & ${M_{B_i}} \cosh(\theta)/{T}$              & $0$                        & $+1$   & $0$                                                                         & $+1$           & $+1$           \\ \hline
Soliton                           & $S$                              & ${M} \cosh(\theta)/{T}$                    & $+1$                       & $+1$   & $0$                                                                         & $0$            & $0$            \\ \hline
First level intermediate magnons  & $m_j$, $j=1,...,\nu_1-1$         & 0                                            & $-2\cdot j$                & $-1$   & $+1$                                                                        & $+1$           & $0$            \\ \hline
First level final magnon          & $m_{\nu_1}$                      & 0                                            & $-2$                       & $+1$   & $-1$                                                                        & $+1$           & $0$            \\ \hline
Second level intermediate magnons & $m_{\nu_1+k}$, $k=1,...,\nu_2-2$ & 0                                            & $-2\cdot(1+k \cdot \nu_1)$ & $+1$   & $(-1)^{\left\lfloor k \cdot \frac{\nu_1\nu_2}{1+\nu_1\nu_2} \right\rfloor}$ & $+1$           & $0$            \\ \hline
Second level next-to-last magnon         & $m_{\nu_1+\nu_2-1}$, $(k=\nu_2-1)$           & $ 2(1+ \nu_1\cdot \nu_2)\cdot{\mu}/{T}$ & $-2\cdot(1+k \cdot \nu_1)$ & $+1$   & $(-1)^{\left\lfloor k \cdot \frac{\nu_1\nu_2}{1+\nu_1\nu_2} \right\rfloor}$ & $+1$           & $0$            \\ \hline
Second level last magnon        & $m_{\nu_1+\nu_2}$                    & $ 2(1+ \nu_1\cdot \nu_2)\cdot{\mu}/{T}$ & $-2\cdot \nu_1$            & $-1$   & $+1$                                                                        & $0$            & $0$            \\ \hline
\end{tabular}
\end{table}

\subsection{Note on higher number of magnonic levels}
The nontrivial ingredient in obtaining the sine-Gordon TBA was to calculate how to ``sew" together the massive and the magnonic parts of the partially decoupled system. When considering more than two magnonic levels, the junction of the massive and the magnonic parts is not affected, so these further TBA systems can be obtained in a straightforward way using the string classification of the XXZ spin chain in the gapless phase \cite{1972PThPh..47...69T,takahashi_1999}. 

\subsection{Cross-checks} 

We tested the validity of our results through a number of cross-checks. First of all, we calculated the free energy density $f$ from the two equivalent expressions 
\begin{align}
        \frac{f}{T} &= \sum_a \int \mathrm{d}\theta \left\{\rho_a^{\text{r}} M_a \cosh\theta - T \left[ \rho_a^{\text{r}} \log\left(1+\frac{\rho_a^{\text{h}}}{\rho_a^{\text{r}}}\right) + \rho_a^{\text{h}} \log\left(1+\frac{\rho_a^{\text{r}}}{\rho_a^{\text{h}}}\right)\right] - \rho_a^{\text{r}} \mu q_a \right\}\,,
\nonumber\\
        \frac{f}{T} &= \sum_a \int \frac{\mathrm{d}\theta}{2\pi} \eta_a M_a\cosh\theta \Big[ -\log(1+\mathrm{e}^{-\epsilon_a})\Big] = \sum_a \int \frac{\mathrm{d}\theta}{2\pi} \eta_a M_a\cosh\theta \big[ -L(\epsilon_a)\big] \,.
        \label{SM:free_energy_for_dressing}
\end{align}
Note that while all the particle species contribute to the first expression, only the massive particles contribute to the second one.

Since we can calculate the densities and the pseudo-energies from the fully coupled (\ref{SM:TBA_coupled_eq}) and the partially decoupled systems (\ref{SM:TBA_eq_decoupled}) (and the corresponding equations for the densities), this gives four different ways of computing the same quantity. We performed extensive cross-checks using all these different relations. Furthermore, taking the derivative of the pseudo-energies as described in Eq. \eqref{eq:density_as_derivative}  results in the densities, which gives another way to cross-check the numerically implemented decoupled density equations \eqref{SM:density_eq}. 

Additionally, an independent verification for the case of vanishing chemical potential can be obtained by comparing the free energies to those computed from the Destri-de Vega (DdV) complex nonlinear integral equation \cite{1995NuPhB.438..413D}
\begin{align}
Z(\theta)&=\frac{M}{T} \sinh \theta 
- i\int\limits_{-\infty}^\infty d\theta' G(\theta-\theta'-i\varepsilon)\log\left( 1+ e^{i Z(\theta'+i\varepsilon)}\right)
+ i\int\limits_{-\infty}^\infty d\theta' G(\theta-\theta'+i\varepsilon)\log\left( 1+ e^{-i Z(\theta'-i\varepsilon)}\right)\,,\nonumber\\
&G(\theta)=\frac{1}{2\pi i} \frac{d}{d\theta}\log S_0(\theta)=\frac{1}{2\pi}
\int\limits_{-\infty}^{\infty} \mathrm{d}t
\frac{\sinh\left(\frac{t\pi}{2}(\xi-1)\right)}{2\sinh\left(\frac{\pi\xi t}{2}\right)\cosh\left(\frac{\pi t}{2}\right)}{e}^{\mathrm{i}\theta t}\,.
\end{align}
The above equation can be solved iteratively for the function $Z(\theta)$, from which the free energy can be computed using the formula
\begin{equation}
    \frac{f}{T}=-2\,\mathrm{Im}\int\limits_{-\infty}^\infty \frac{\mathrm{d}\theta}{2\pi} M\sinh(\theta+i\varepsilon)\log\left( 1+ e^{i Z(\theta+i\varepsilon)}\right)\,.
\end{equation}
This comparison was performed for various values of $\xi$ and $T$, with a sample of the results shown in Table \ref{tab:tbaddv}.

\begin{table}[h]
\centering
\begin{tabular}{|c|cc|cc|cc|}
\hline
$T/M$ & \multicolumn{2}{c|}{$\xi=1/3$}                           & \multicolumn{2}{c|}{$\xi=1/(1+\frac{1}{3})$}             & \multicolumn{2}{c|}{$\xi=3$}                             \\ \hline\hline
      & \multicolumn{1}{c|}{DdV}              & TBA              & \multicolumn{1}{c|}{DdV}              & TBA              & \multicolumn{1}{c|}{DdV}              & TBA              \\ \hline
$20$  & \multicolumn{1}{c|}{$-10.4648558878$} & $-10.4648558858$ & \multicolumn{1}{c|}{$-10.4512155761$} & $-10.4512155756$ & \multicolumn{1}{c|}{$-10.3511969646$} & $-10.3511979842$ \\ \hline
$10$  & \multicolumn{1}{c|}{$-5.22194300631$} & $-5.22194300532$ & \multicolumn{1}{c|}{$-5.19859206357$} & $-5.19859206333$ & \multicolumn{1}{c|}{$-5.10437699605$} & $-5.10437749589$ \\ \hline
$5$   & \multicolumn{1}{c|}{$-2.59068229582$} & $-2.59068229533$ & \multicolumn{1}{c|}{$-2.55324704632$} & $-2.55324704620$ & \multicolumn{1}{c|}{$-2.47030920896$} & $-2.47030944780$ \\ \hline
$2$   & \multicolumn{1}{c|}{$-0.98474359521$} & $-0.98474359502$ & \multicolumn{1}{c|}{$-0.92684324811$} & $-0.92684324807$ & \multicolumn{1}{c|}{$-0.87159594008$} & $-0.87159602002$ \\ \hline
$1$   & \multicolumn{1}{c|}{$-0.41783483396$} & $-0.41783483387$ & \multicolumn{1}{c|}{$-0.35869323482$} & $-0.35869323480$ & \multicolumn{1}{c|}{$-0.33218987940$} & $-0.33218990576$ \\ \hline
$0.5$ & \multicolumn{1}{c|}{$-0.11948181719$} & $-0.11948181716$ & \multicolumn{1}{c|}{$-0.08847607829$} & $-0.08847607828$ & \multicolumn{1}{c|}{$-0.08339026939$} & $-0.08339027349$ \\ \hline
$0.2$ & \multicolumn{1}{c|}{$-0.00386984298$} & $-0.00386984298$ & \multicolumn{1}{c|}{$-0.00258436645$} & $-0.00258436645$ & \multicolumn{1}{c|}{$-0.00256472781$} & $-0.00256472782$ \\ \hline
\end{tabular}
\caption{\label{tab:tbaddv} Comparison of the free energy $f/T$ computed from the DdV and the TBA approaches at a reflectionless, an attractive, and a repulsive coupling. Depending on the coupling the relative errors between the two results are between $10^{-4}-10^{-8}$, and it gets better for smaller tolerance/more iteration.}
\end{table}

\section{Dressing relation and effective velocity}

Following  \cite{Doyon_2020} to derive the dressing equations, we write the source terms in Eq.~(\ref{SM:TBA_eq_decoupled}) as
\begin{equation}
    w_a = \sum_b \beta^b Q_a^{(b)}(\theta) = \frac{1}{T}M_a\cosh\theta + \frac{\lambda}{T} M_a \sinh\theta - \frac{\mu}{T} q_a\,,
    \label{SM:GGE_source}
\end{equation}
where $Q_a^{(e)}=M_a\cosh \theta$, $Q_a^{(p)}=M_a\sinh\theta$ and $Q_a^{(q)}=q_a$ and $1/T$, $\lambda/T$, $\mu/T$, are the charges (of particle $a$) and conjugate temperature variables associated to the energy, momentum and topological charge. We note that this idea can be extended to construct generalised Gibbs ensembles from TBA \cite{Mossel_2012}.

Starting from the free energy (\ref{SM:free_energy_for_dressing}), the expectation value of a charge conjugate to the generalised temperature variables $\beta^k$
\begin{equation}
    \textrm{q}_k = \frac{\partial}{\partial \beta^k}\frac{f}{T} = \sum_a \int \frac{\mathrm{d}\theta}{2\pi} M_a\cosh\theta \left(\frac{\partial (-L)}{\partial \epsilon_a}\right)\left(\frac{\partial \epsilon_a}{\partial \beta^k}\right)\eta_a = \sum_a \int \frac{\mathrm{d}\theta}{2\pi} M_a\cosh\theta\,\vartheta_a \left(Q_a^{(k)}\right)^{\textrm{dr}}\,.
\end{equation}
Using Eqs.(\ref{SM:TBA_eq_decoupled},\ref{SM:GGE_source}), and omitting the superscript describing the type of the charge to lighten the notation, the dressed charge satisfies the dressing equation
\begin{equation}
    \eta_a\,Q_a^{\text{dr}} =  Q_a + \sum_j K_{ab}*\left[\left(\sigma_b^{(1)}-\vartheta_b\right)\eta_b\,Q_b^{\text{dr}}-\sigma_b^{(2)} Q_b \right]\,.
    \label{SM:dressing}
\end{equation}
Note that in our calculations we used $\lambda=0$, $\mu=0$, corresponding to the centre-of-mass frame and zero topological charges, i.e. we needed $\lambda$ and $\mu$ to derive the dressing equations but otherwise set them to zero.

\begin{figure}[t]
\centering
\captionsetup[subfigure]{justification=centering}
\begin{subfigure}{0.32\textwidth}
    \includegraphics[width=1\textwidth]{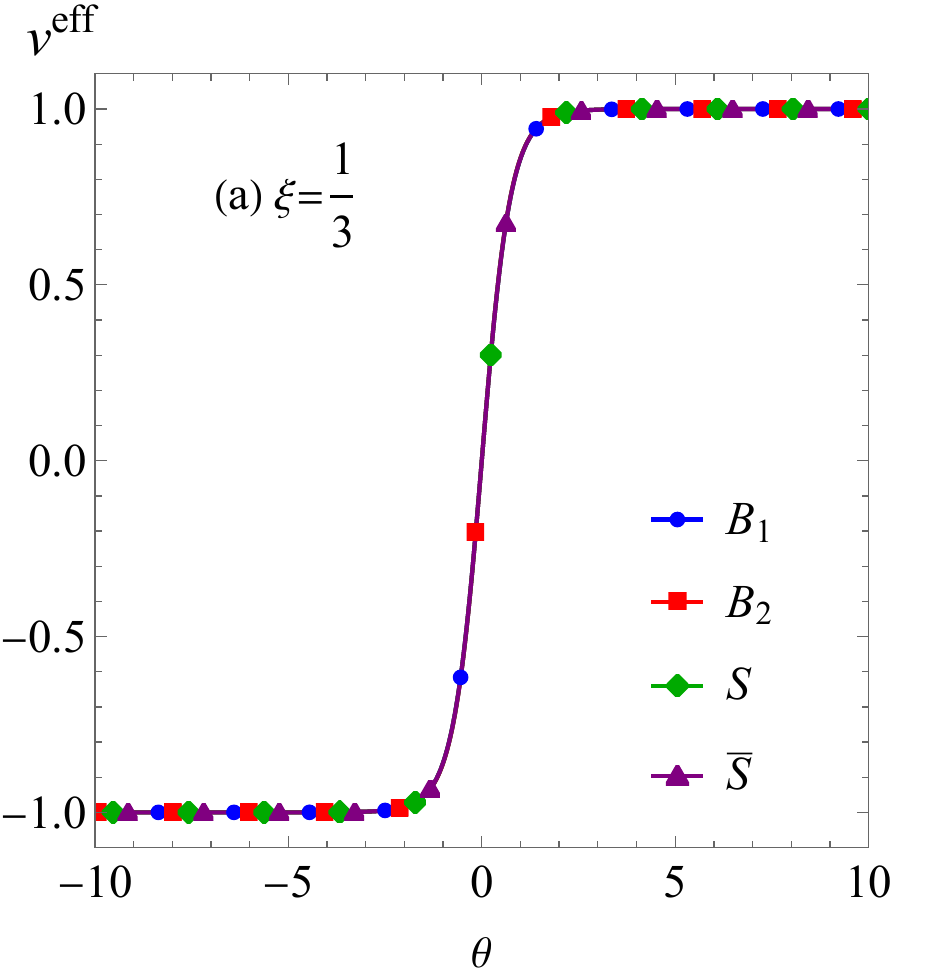}
\end{subfigure}
\hfill
\begin{subfigure}{0.32\textwidth}
    \includegraphics[width=1\textwidth]{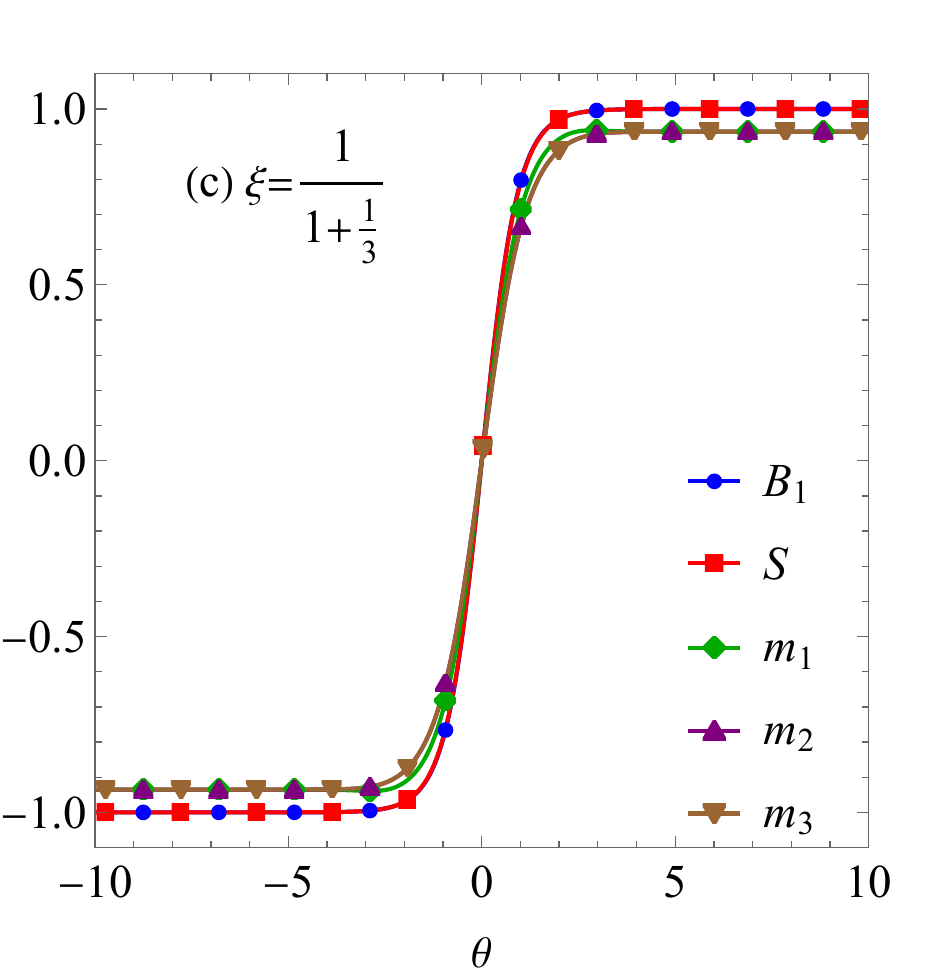}
\end{subfigure}
\hfill
\begin{subfigure}{0.32\textwidth}
    \includegraphics[width=1\textwidth]{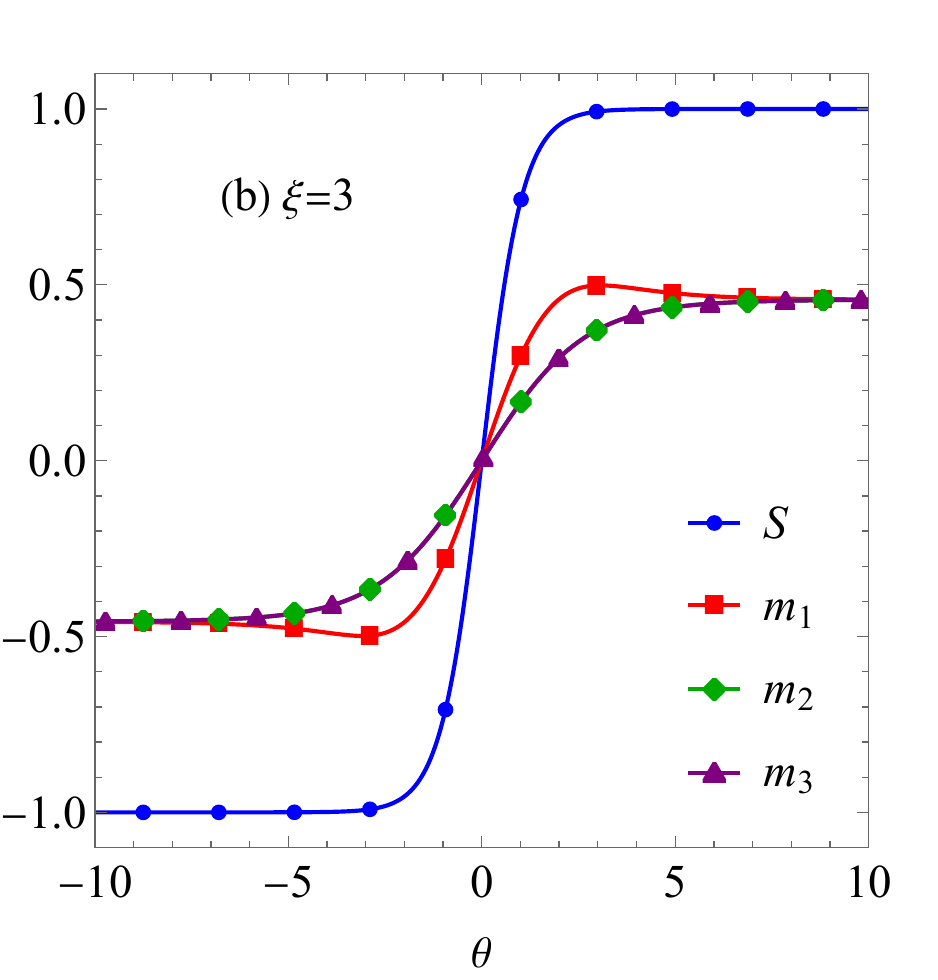}
\end{subfigure}
\caption{Effective velocity at $T=1$ at three different couplings. Note that the effective velocity of magnons gets close to one in the attractive regime, but has a very different shape in the repulsive regime, and in particular has a maximum that is significantly less than one.}
\label{SM:eff_vel}
\end{figure}

The propagation velocity of excitations is modified by interactions which in the hydrodynamic description is taken into account by the effective velocity
\begin{equation}
    v^{\text{eff}}_a(\theta) = \frac{(\partial_{\theta}e_a)^{\text{dr}}(\theta)}{(\partial_{\theta}p_a)^{\text{dr}}(\theta)} = \frac{(\partial_{\theta}e_a)^{\text{dr}}(\theta)}{2\pi \rho^{\text{tot}}_a(\theta)}\,,
\end{equation}
with a few examples shown in Fig.~\ref{SM:eff_vel}.

\section{Drude weight from the bipartitioning protocol}

One of the most common non-equilibrium setups to study transport in quantum systems is the bipartitioning protocol (or Riemann problem) when the system is divided into two semi-infinite lines, with the initial states in the two parts chosen as local equilibrium states corresponding to the source terms
\begin{align}
    w_a = 
    \begin{cases}
        w_{a,L} &= \sum_b \beta^b_L Q_a^{(b)}\,,\quad x<0\,, \\
        w_{a,R} &= \sum_b \beta^b_R Q_a^{(b)}\,,\quad x>0\,.
    \end{cases}
\end{align}
The asymptotic stationary value of the filling fraction along a line with fixed $\zeta=x/t$ can be computed as \cite{2016PhRvL.117t7201B,2016PhRvX...6d1065C}
\begin{equation}
    \vartheta_a(\zeta, \theta) = \Theta\left(v_a^{\text{eff}}(\zeta,\theta)-\zeta\right)\vartheta_{a,L}(\theta) + \Theta\left(\zeta-v_a^{\text{eff}}(\zeta,\theta)\right)\vartheta_{a,R}(\theta)\,,
    \label{SM:bip_filling}
\end{equation}
where $\Theta$ is the Heaviside step function, and $\vartheta_{a,L/R}$ are the filling fractions of the asymptotic states of the left and the right side of the bipartite system. Note that Eq.~(\ref{SM:bip_filling}) is only an implicit equation for $\vartheta_a$, because the effective velocity depends on the filling fractions, but a recursive numerical scheme quickly converges to the solution as described in e.g. \cite{Doyon_2020,2019PhRvB.100c5108B}.

\begin{figure}[t]
\centering
\captionsetup[subfigure]{justification=centering}
\begin{subfigure}{0.32\textwidth}
    \includegraphics[width=1\textwidth]{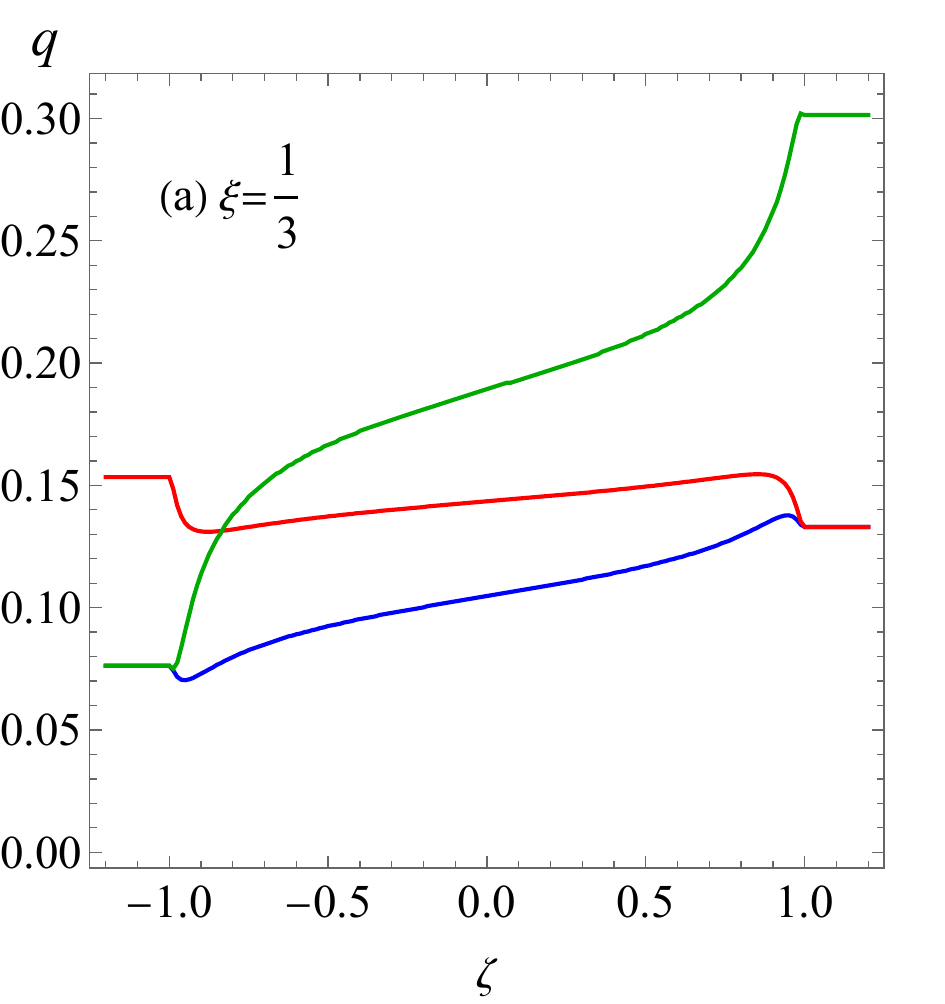}
\end{subfigure}
    \hfill
\begin{subfigure}{0.32\textwidth}
    \includegraphics[width=1\textwidth]{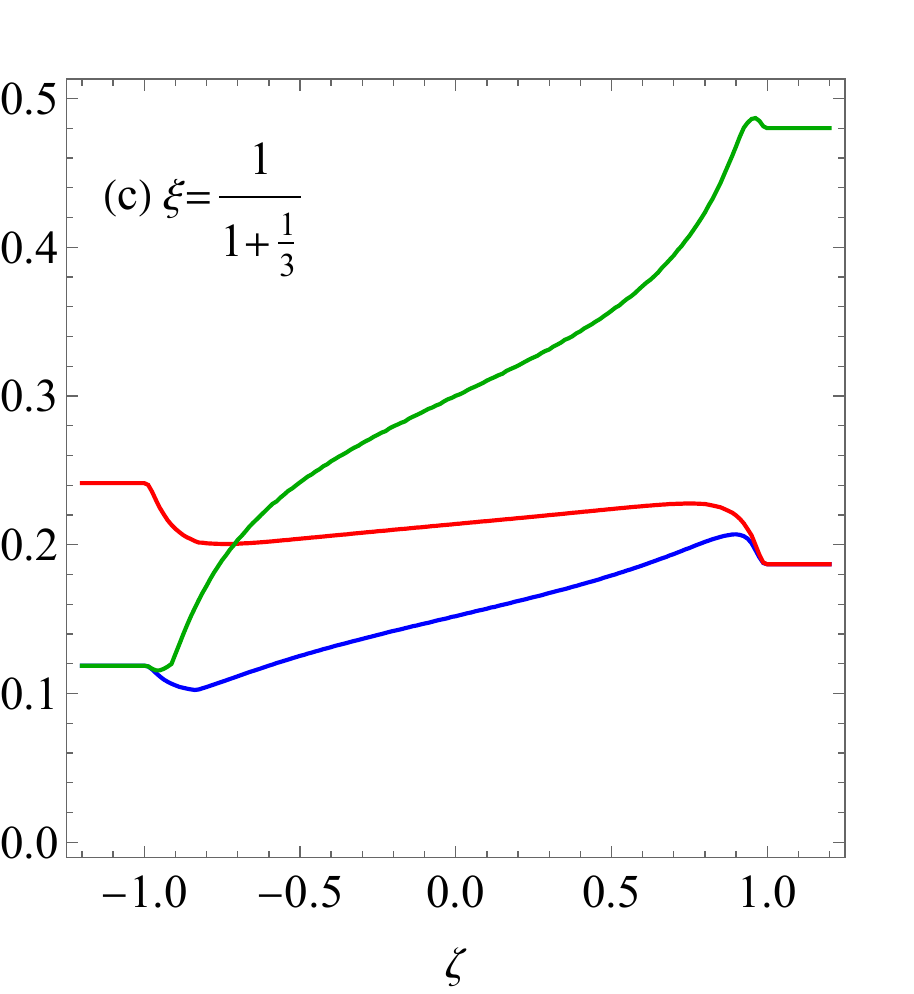}
\end{subfigure}
\hfill
\begin{subfigure}{0.32\textwidth}
    \includegraphics[width=1\textwidth]{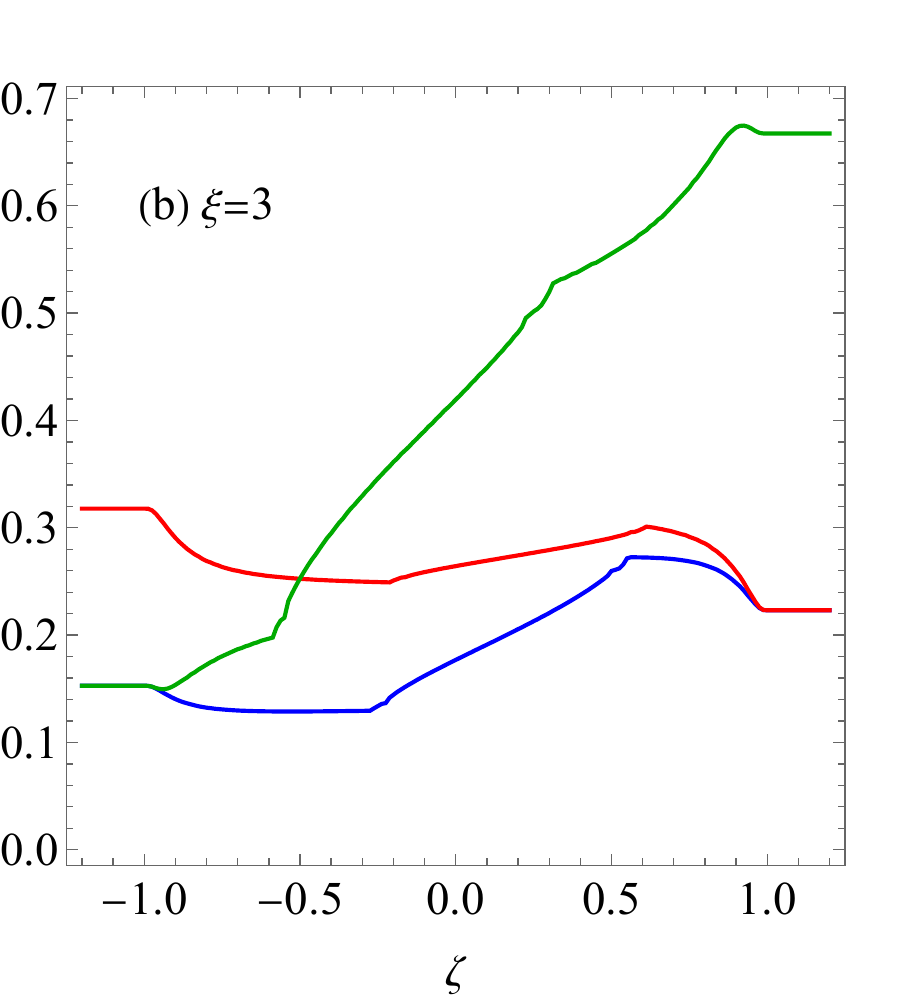}
\end{subfigure}
\begin{subfigure}{0.32\textwidth}
    \includegraphics[width=1\textwidth]{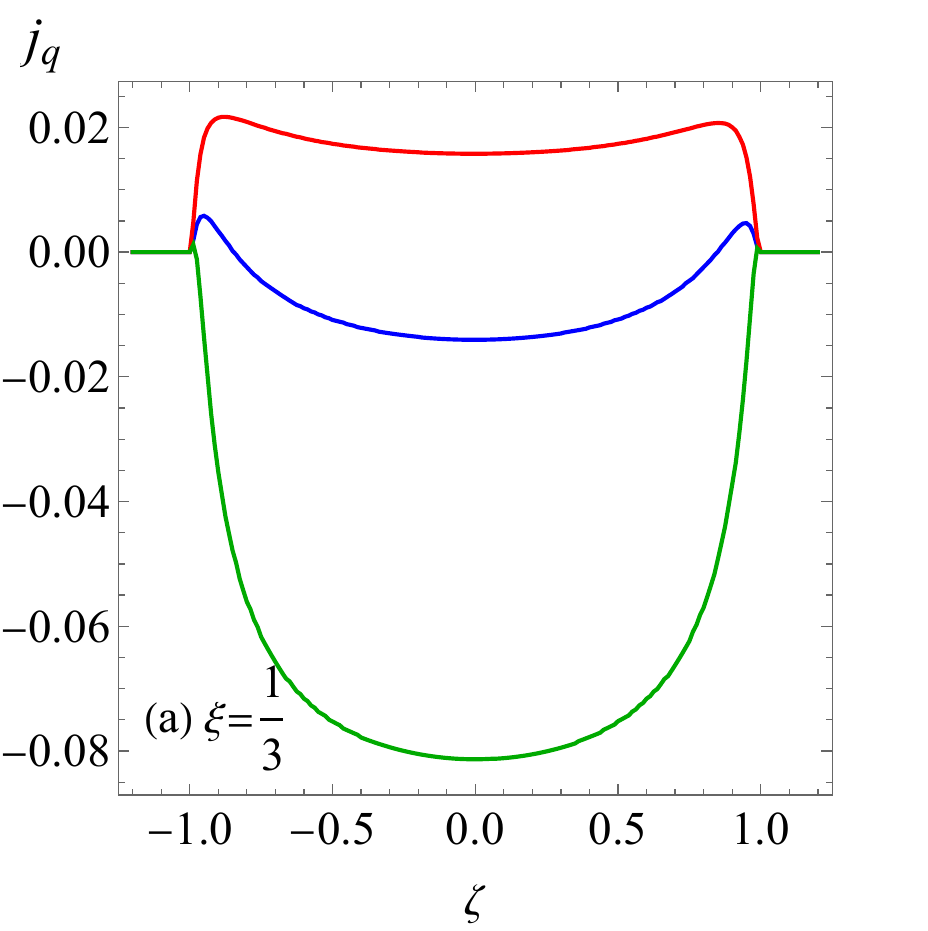}
\end{subfigure}
\hfill
\begin{subfigure}{0.32\textwidth}
    \includegraphics[width=1\textwidth]{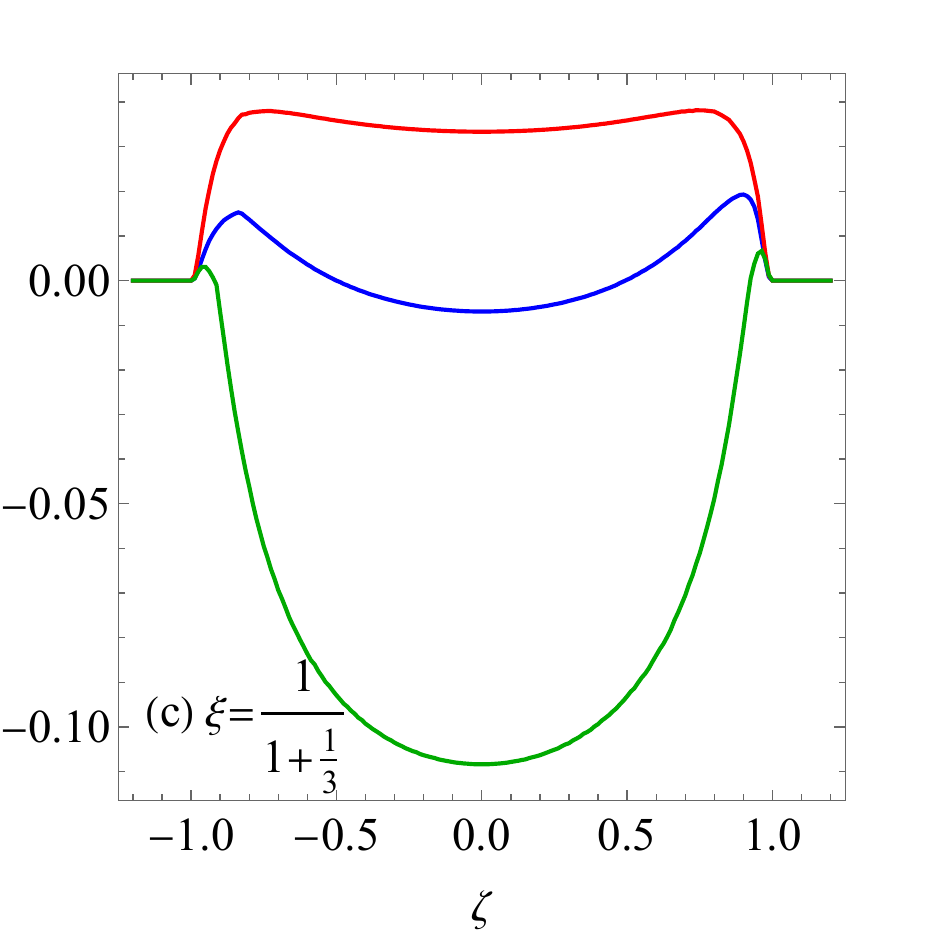}
\end{subfigure}
\hfill
\begin{subfigure}{0.32\textwidth}
    \includegraphics[width=1\textwidth]{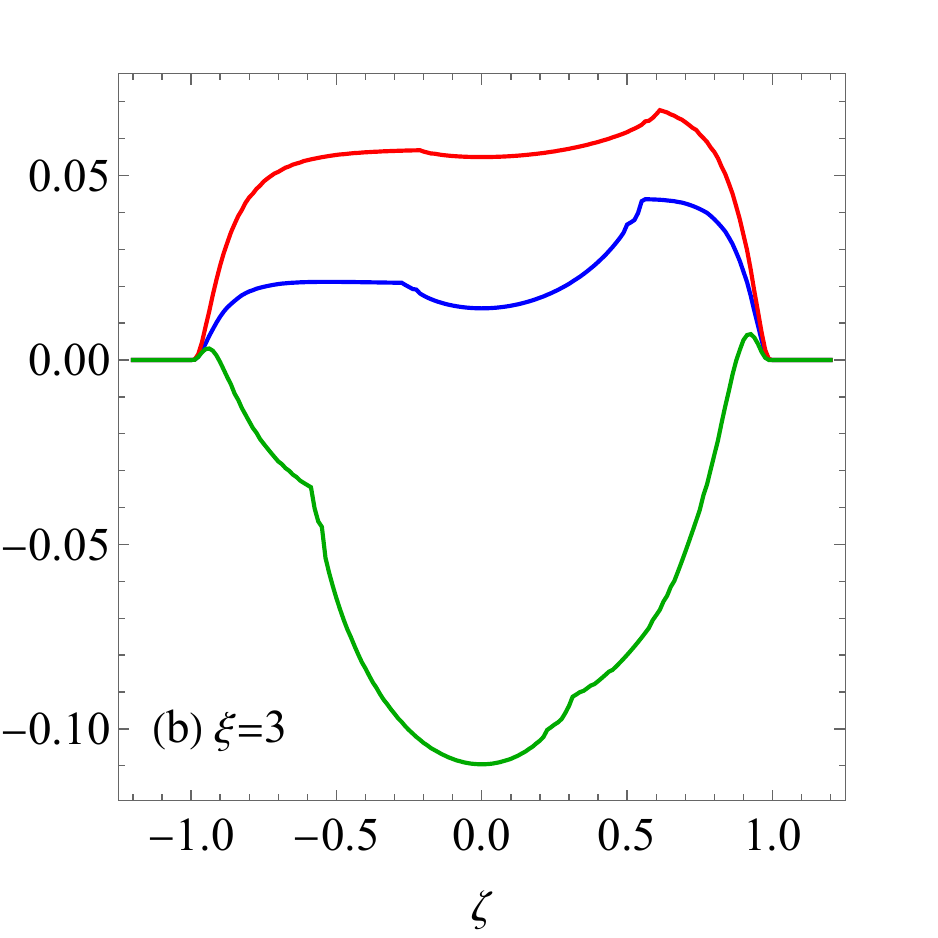}
\end{subfigure}
\caption{$T_L=1\,M$, $T_R=0.5\,M$, blue: $\mu_L=0.5\,M$, $\mu_R=1\,M$, red: $\mu_L=1\,M$, $\mu_R=1\,M$, green: $\mu_L=0.5\,M$, $\mu_R=2\,M$. Note the visible discontinuities in the profiles in the repulsive regime, caused by the nontrivial maximum of the effective velocities of the magnons \cite{Piroli_2017}. In the attractive regime, the discontinuities are not that pronounced, because the maximal magnon velocities are close to one as shown in Fig. \ref{SM:eff_vel}. However, Fig. \ref{SM:eff_vel} is only indicative, as there is a different set of effective velocities for all rays in the partitioning protocol.}
\label{SM:bip_figs}
\end{figure}

Once the filling fractions are known, one can calculate the expectation values of the topological charge and its current as
\begin{align}
    \text{q}(\zeta) &= \sum_a \int \mathrm{d}\theta \rho_a^{\textrm{tot}}(\zeta,\theta) \vartheta_a(\zeta,\theta) q_a 
    = \sum_a \int \mathrm{d}\theta \frac{M_a}{2\pi} \cosh(\theta) \vartheta_a(\zeta,\theta) q_a^{\textrm{dr}}(\zeta,\theta)\,,\nonumber \\
    \text{j}(\zeta) &= \sum_a \int \mathrm{d}\theta \rho_a^{\textrm{tot}}(\zeta,\theta) 
    \vartheta_a(\zeta,\theta) q_a v^{\textrm{eff}}(\zeta,\theta)
    = \sum_a \int \frac{\mathrm{d}\theta}{2\pi} (\epsilon_a')^{\textrm{dr}}(\zeta,\theta) \vartheta_a(\zeta,\theta) q_a 
    = \sum_a \int \frac{\mathrm{d}\theta}{2\pi} M_a \sinh(\theta) \vartheta_a(\zeta,\theta) q_a^{\textrm{dr}}(\zeta,\theta)\,,
\end{align}
with example profiles shown in Fig. \ref{SM:bip_figs}.

The Drude weight of the topological charge is defined as
\begin{equation}
D_q={\lim_{\tau\to\infty}} \frac{1}{2\tau}\int\limits_{-\tau}^{\tau} \mathrm{d}t \int \mathrm{d}x \,\langle j_q(x,t) j_q(0,0)\rangle^c\,,
\label{SM:Drude_weight_def}
\end{equation}
and can be computed from the bipartitioning protocol by preparing the two halves of the system with different chemical potentials and integrating over the current \cite{Doyon_2017} as
\begin{equation}
    D_q = \left.\frac{\partial}{\partial \delta \mu} \int \mathrm{d}\zeta\ \textrm{j}\left(\zeta\right)\right|_{\delta\mu=0}\,.
    \label{Drude_bipartition}
\end{equation}
We found that choosing $\delta \mu=5\cdot 10^{-3}$ reproduces the result calculated from (\ref{Drude_TBA}) within $0.1\%$ relative difference, further validating the consistency of our methods and the TBA system (\ref{SM:TBA_eq_decoupled}).

\section{Limiting cases}

\begin{figure}[t]
\centering
\captionsetup[subfigure]{justification=centering}
\begin{subfigure}{0.49\textwidth}
    \centering
    \includegraphics[height=5.8cm]{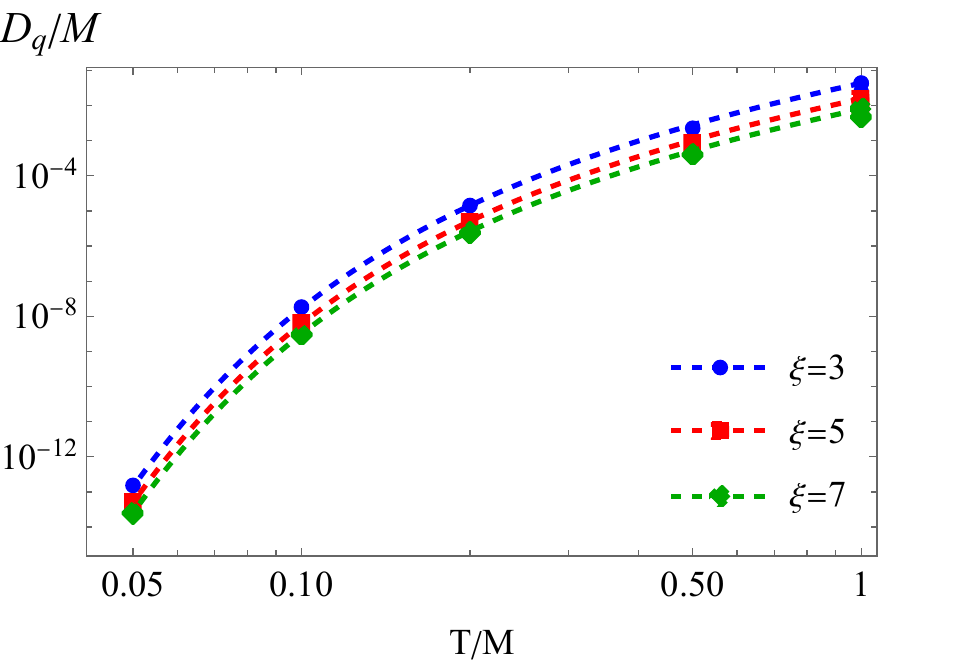}
    \end{subfigure}
    \hfill
\begin{subfigure}{0.49\textwidth}
    \centering
    \includegraphics[height=5.8cm]{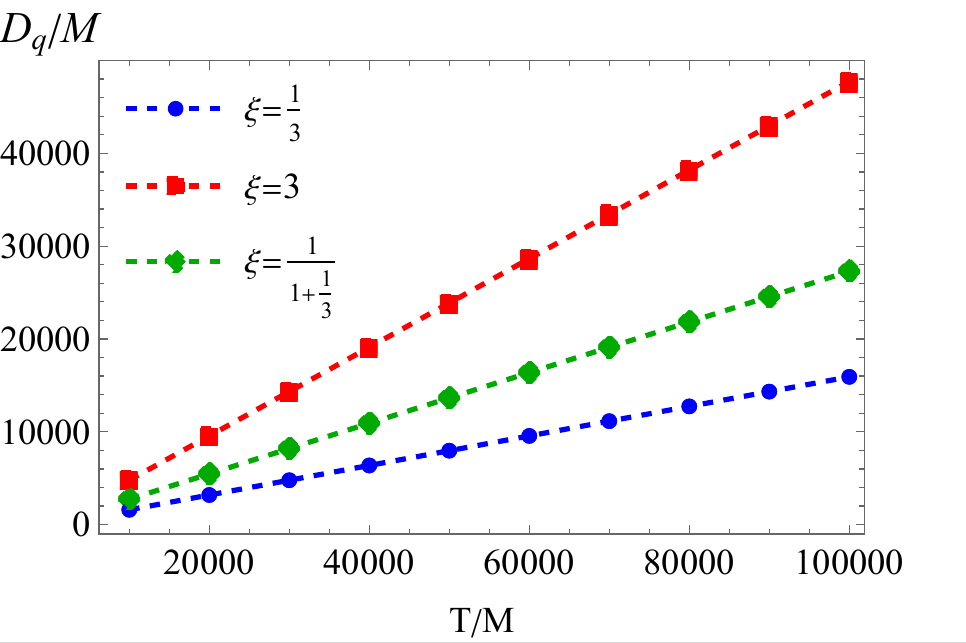}
\end{subfigure}
\caption{Drude weights as a function of temperature. The dashed lines in the left and right panel are the analytic expressions Eqs.(\ref{low_T_Drude}) and (\ref{SM:highT_lim}), showing excellent agreement with the numerical data.}
\label{SM:D_in_T}
\end{figure}

The Drude weight \eqref{SM:Drude_weight_def} can also be obtained from the TBA using \cite{2019ScPP....6....5U,2022JSMTE2022a4002D}
\begin{equation}
    D_q = \sum_a\!\int\!\mathrm{d}\theta \rho_a^{\textrm{tot}}(\theta) \vartheta_a(\theta)[1-\vartheta_a(\theta)]\left[v_a^{\textrm{eff}}(\theta)q_a^{\textrm{dr}}(\theta)\right]^2\,.
    \label{Drude_TBA}
\end{equation}
Below we compute its low- and high-temperature behaviour analytically, which can be used to cross-checks our numerical results as shown in Fig.~\ref{SM:D_in_T}. 

\subsection{Low-temperature limit}

In the limit $T\ll M,M_{B_i},$ the TBA equations simplify considerably. For the massive nodes, the source terms dominate, so the $e^{\epsilon(\th)}$ functions become exponentially large, $\sim e^{M\cosh\th}$. As a consequence, the convolution terms expressing coupling to the other massive particles can be neglected. The pseudo-energies of the massless magnons remain finite, so the coupling of the soliton to the first magnon needs to be kept. 

Since $e^{\epsilon_S(\th)}$ becomes exponentially large, it can be neglected in the  equation for $\epsilon_{m_1}(\th)$. Then the magnonic TBA equations completely decouple from the kink equation. In a thermal state, the source terms are constant, so the pseudo-energies take  constant values $\tilde{\epsilon}_{m_k}$, and their exponentials $y_k = \exp(\tilde{\epsilon}_{m_k})$ satisfy algebraic equations. For one magnonic level they read
\begin{align}
y_1^2 &= 1+y_2\,,\nonumber\\
y_n^2 &= (1+y_{n-1})(1+y_{n+1})\,,\qquad\qquad\qquad  1<n<p-2\,,\nonumber\\
y_{p-2}^2 &= (1+y_{p-3})(1+y_{p-1})(1+y_p^{-1})\,,\nonumber\\
y_{p-1}^2 &= e^{2p\mu/T}(1+y_{p-2})\,,\nonumber\\
y_{p}^2 &= e^{2p\mu/T}(1+y_{p-2})^{-1}\,,
\end{align}
where we used that the integral of the kernel over $2\pi$ is equal to $1/2.$ 
The solution for $\mu=0$ is
\begin{align}
&y_{m_k} = (k+1)^2-1\,,\qquad k=1,\dots,p-2\,,\nonumber\\
&y_{m_{p-1}}= 1/y_{m_p} = p-1\,.
\label{eq:lowTmag}
\end{align}
For the soliton we have 
$
y_S(\theta) = e^{M\cosh(\theta)/T}/2,
$
while for the breathers 
$
y_{B_i}(\theta) = e^{M_{B_i}\cosh(\theta)/T}.
$

The root densities and dressed velocities can also be obtained in a closed form. For the massive nodes,
\be
\rho_S(\th)=2\frac{M\cosh\th}{2\pi} e^{-M\cosh\th/T}\,,\qquad \rho_{B_i}(\th)=\frac{M_{B_i}\cosh\th}{2\pi} e^{-M\cosh\th/T}\,,
\ee
while for the magnons \cite{Bertini:2019lzy}
\begin{align}
\rho_{m_k}(\theta) &= \frac{n}{4\pi} \frac1{k+1}\left(\frac{a_k(\theta)}{k}-\frac{a_{k+2}(\theta)}{k+2}\right)\,,\qquad k=1,\dots,p-2\,,\nonumber\\ 
\rho_{m_{p-1}}(\theta) &= \frac{n}{4\pi} \frac{a_{p-1}(\theta)}{p-1}\,,\\
\rho_{m_p}(\theta) &= \frac{n}{4\pi} a_{p-1}(\theta)\,,
\label{rholimmu0}
\end{align}
where 
\be
\label{eq:n}
n=\int\mathrm{d}\th \rho_S(\th) \approx \sqrt{2MT/\pi}e^{-M/T}
\ee
is the kink density, and
\be
a_k(\th) = \frac2p \frac{\sin\left(\frac{k\pi}p\right)}{\cosh\left(\frac{2\th}p\right)-\cos\left(\frac{k\pi}p\right)}\,.
\ee
The TBA equations \eqref{SM:TBA_eq_decoupled} imply that all filling fractions of massive excitations are exponentially small at low temperatures, implying that the dressed velocity of the massive particles is equal to their bare velocity $v(\th)=\tanh\th$, with similar precision. In this limit, the dressed velocities of the magnons can be evaluated as
\be
\label{veff_lowT}
v_{m_k}(\th) \approx -\frac{T}M \frac{\rho'_k(\th)}{\rho_k(\th)}\,.
\ee
At $\mu=0$ only the last two magnons carry dressed topological charge $p$, so the expression \eqref{Drude_TBA} for the Drude weight yields 
\be
D_q = 2 \int\mathrm{d}\th \rho_{m_p}(\th)(1-\tilde\vartheta_{m_p})[v_{m_p}(\th)p]^2 \approx 
\frac{n}{2\pi}\frac{T^2}{M^2}p\int\mathrm{d}\theta\frac{a_{p-1}'(\th)^2}{a_{p-1}(\th)}
=\frac{n}\pi \frac{T^2}{M^2} \frac{2\pi/p - \sin(2\pi/p)}{p\sin^2(\pi/p)}\,,
\ee
with the complete result coming exclusively from the last magnons, which contribute equally hence the factor $2$. Using Eq. \eqref{eq:n}, the low-temperature behaviour of the Drude weight for large $p$ is given by 
\begin{equation}
    D_q = \frac{\sqrt{2} T^{5/2}}{\sqrt{\pi}M^{3/2}}\frac{4}{3 p^2}{e}^{-M/T}  \, .
    \label{low_T_Drude}
\end{equation} 

\subsection{High-temperature limit}

In the limit $T\gg M,M_{B_a},$ the source terms of the TBA equations of the massive particles of mass $M_a$ can be neglected in the region $-\log(2T/M_a)\lesssim \th \lesssim \log(2T/M_a).$ As a consequence, the pseudo-energies in this region become constant, and their corresponding ``plateau'' values satisfy a set of algebraic equations that for one magnonic level in the generic case reads
\begin{align}
\bar{y}_{B_1}^2 &= 1+\bar{y}_{B_2}\,,\nonumber\\
\bar{y}_{B_l}^2 &= (1+\bar{y}_{B_{l-1}})(1+\bar{y}_{B_{l+1}})\,,  &&1<l<n_B\,,\nonumber\\
\bar{y}_{B_l}^2 &= (1+\bar{y}_{B_{l-1}})(1+\bar{y}_{B_l})(1+\bar{y}_S)\,,  &&l=n_B\,,\nonumber\\
\bar{y}_S^2 &= (1+\bar{y}_{B_l})/(1+\bar{y}_{m_1})\,,  &&l=n_B\,,\nonumber\\
\bar{y}_{m_k}^2 &= (1+\bar{y}_{m_{k-1}})(1+\bar{y}_{m_{k+1}})\,,  &&1<k<p-2\,,\nonumber\\
\bar{y}_{m_{p-2}}^2 &=  (1+\bar{y}_{m_{p-3}})(1+\bar{y}_{m_{p-1}})(1+\bar{y}_{m_p}^{-1})\,,\nonumber\\
\bar{y}_{m_{p-1}}^2 &= e^{2p\mu/T}(1+\bar{y}_{m_{p-2}})\,,\nonumber\\
\bar{y}_{m_p}^2 &= e^{2p\mu/T}(1+\bar{y}_{m_{p-2}})^{-1}\,.
\end{align}
For the cases of one magnonic level and $\mu=0$ the solution is
\begin{align}
\bar{y}_{B_k} &= (k+1)^2-1\,,\nonumber\\
\bar{y}_S &= \left[\left(\frac{n_B+2}{n_B+1}\right)^2-1\right]^{-1}\,,\nonumber\\
\bar{y}_{m_k} & = \left(k+\frac{n_B+2}{n_B+1}\right)^2-1\,,\qquad \qquad k=1,\dots,p-2\,,\nonumber\\
\bar{y}_{m_{p-1}}&= \bar{y}_{m_p}^{-1} = p-\frac{n_B}{n_B+1}\,.
\label{eq:plateaux}
\end{align}
These equations hold also in the attractive regime with $n_B=0.$ In the reflectionless case there are no magnons but we have two solitonic nodes. The breather plateau values are the same, while for the soliton and antisoliton $\bar{y}_S = \bar{y}_{\bar S} = n_B+1$.

In the region $|\th|\gg \log 2T/M_a,$ the pseudo-energy functions for the massive excitations diverge as $\sim M_a\cosh\th$, so their filling functions rapidly decay to zero, while the magnonic pseudo-energies take their constant low-temperature values \eqref{eq:lowTmag} as $|\th|\to\infty.$ The root densities $\rho_a(\theta)$ are peaked around $\th= \log 2T/M_a$ and otherwise exponentially close to zero, so the dressing equations decouple into independent left and right moving modes. In the peak regions, the effective velocities of all the excitations are $\pm1.$

The free energy density can be expressed using the standard dilogarithm trick \cite{Andrews:1984af,Zamolodchikov:1989cf,Klassen:1990dx}, with the use of Roger's dilogarithm function 
\be
L(x) = -\frac12\int\limits_0^x\mathrm{d} y \left(\frac{\log y}{1-y} + \frac{\log(1-y)}y\right)
\ee
as
\be
f = T^2\left(\sum_{k=1}^{n_B} L(\bar\vartheta_{B_k}) + L(\bar\vartheta_S) - 
\sum_{j=1}^{n_m}\eta_{m_j}\left[L(\bar\vartheta_{m_j})-L(\tilde\vartheta_{m_j})\right]\right)\,,
\ee
where the constant filling fractions 
\begin{equation}
    \tilde\vartheta_a=\frac{1}{1+y_a}\,,\quad \bar\vartheta_a=\frac{1}{1+\bar{y}_a}
\end{equation} 
in terms of the corresponding plateau values. Substituting the solutions of Eqs. \eqref{eq:lowTmag} and \eqref{eq:plateaux}, we check numerically in several cases that 
\be
f=\frac{\pi^2T^2}6
\ee
equal to the exact result for free massless bosons, which is  another consistency check of the validity of our TBA equations. From a mathematical point of view, this result corresponds to nontrivial dilogarithm identities \cite{1995PhLB..355..157T,2012arXiv1212.6853N}.

For the Drude weight we use an additional relation valid at high temperature \cite{Klassen:1990dx}:
\be
\rho_a(\th) = -\frac{T}{2\pi}\frac{\mathrm{d}}{\mathrm{d}\th} \log\left(1+e^{-\epsilon_a(\th)}\right) = \frac{T}{2\pi} \epsilon_a'(\th)\vartheta_a(\th)\,,
\ee
which can be plugged into the formula \eqref{Drude_TBA} with the result
\be
D_q = 2 \frac{T}{2\pi}\int\mathrm{d}\th \epsilon_{p-1}'(\th)\vartheta_{p-1}(\th)[1-\vartheta_{p-1}(\th)]p^2\,,
\ee
where we used that at very high temperatures all velocities are equal to $\pm 1$. The integrand turns out to be a total derivative, so
\be
D_q = \frac{T}\pi p^2 \,2\int\limits_0^\infty\mathrm{d}\th \vartheta_{p-1}'(\th) = \frac{2T}\pi p^2 ( \tilde\vartheta_{p-1} - \bar\vartheta_{p-1} )
= \frac{2T}\pi p^2 \left(\frac1p - \frac1{1+p-\frac{n_B}{n_B+1}}\right)
= \frac{2T}\pi \frac\xi{\xi+1}\,.
\label{SM:highT_lim}
\ee
Below we show that this result is independent of the magnon structure. As a result, the high-temperature limit of the Drude weight is a continuous function of the coupling parameter $\xi$, and the fractal structure is suppressed.

\subsection{High-temperature limit from free boson theory}

In the high-temperature limit, the sine-Gordon interaction can be neglected and the Hamiltonian can be approximated by a massless free boson. We start by considering a free massive boson in a finite volume $L$
\begin{align}
\phi(t,x) & =\frac{1}{\sqrt{L}}\sum_{k}\frac{1}{\sqrt{2\omega_{k}}}\left(a_{k}e^{-i\omega_{k}t+ikx}+a_{k}^{\dagger}e^{i\omega_{k}t-ikx}\right)
\end{align}
with the dispersion relation $\omega_k=\sqrt{k^2+m^2}$, and the corresponding Hamiltonian is
\begin{align}
H & =\frac{1}{2}\int dx:\left(\partial_{t}\phi\right)^{2}+\left(\partial_{x}\phi\right)^{2}+m^{2}\phi^{2}:\,=\sum_{k}\omega_{k}a_{k}^{\dagger}a_{k}\,.
\end{align}
The finite temperature correlator can be evaluated by elementary manipulations:
\begin{equation}
\left\langle \partial_{t}\phi(t,x)\partial_{t}\phi(0,0)\right\rangle _{T}=\frac{1}{L}\sum_{k}\frac{\omega_{k}}{2}\left(\frac{e^{\omega_{k}/T}}{e^{\omega_{k}/T}-1}e^{-i\omega_{k}t+ikx}+\frac{1}{e^{\omega_{k}/T}-1}e^{i\omega_{k}t-ikx}\right)\,.
\end{equation}
In the massless limit $m\rightarrow 0$, it becomes
\begin{equation}
\left\langle \partial_{t}\phi(t,x)\partial_{t}\phi(0,0)\right\rangle _{T}=\frac{1}{L}\sum_{k\neq0}\frac{|k|}{2}\left(\frac{e^{|k|/T}}{e^{|k|/T}-1}e^{-i|k|t+ikx}+\frac{1}{e^{|k|/T}-1}e^{i|k|t-ikx}\right)+\frac{T}{L}
\end{equation}
which can be integrated to give 
\begin{equation}
\lim_{\tau\rightarrow\infty}\frac{1}{2\tau}\int\limits_{-\tau}^{\tau}dt\int\limits_0^L dx\left\langle \partial_{t}\phi(t,x)\partial_{t}\phi(0,0)\right\rangle _{T}=T\,.
\end{equation}
Using the expression $j_{q}=-\beta\partial_{t}\phi/{2\pi}$ for the topological current then gives 
\begin{equation}
D_{q}=\lim_{\tau\rightarrow\infty}\frac{1}{2\tau}\int\limits_{-\tau}^{\tau}dt\int dx\left\langle j_q(t,x) j_q(0,0)\right\rangle _{T}=T\frac{\beta^{2}}{4\pi^{2}}=\frac{2 T}{\pi} \frac{\xi}{\xi + 1}\,.
\end{equation}

\bibliography{drude}